\documentclass[a4paper,UKenglish]{lipics-v2018}

\usepackage{amssymb,amsmath}
\usepackage[table]{xcolor}
\usepackage{multirow}
\usepackage{tikz}
\usepackage{empheq}
\usepackage{booktabs}
\usepackage[symbol]{footmisc}

\nolinenumbers

\usetikzlibrary{matrix}

\DeclareMathAlphabet\mathbfcal{OMS}{cmsy}{b}{n}

\newcommand{\logic}{{\textsf{EQSMT}}~}
\newcommand{\logicWithPeriod}{{\textsf{EQSMT.}}~}
\newcommand{\logicWithComma}{{\textsf{EQSMT,}}~}

\newcommand{\Ss}{\mathcal{S}}
\newcommand{\Ff}{\mathcal{F}}
\newcommand{\Rr}{\mathcal{R}}
\newcommand{\Rf}{\mathfrak{R}}
\newcommand{\Aa}{\mathcal{A}}
\newcommand{\Mm}{\mathcal{M}}
\newcommand{\Uu}{\mathcal{U}}
\newcommand{\Ii}{\mathcal{I}}
\newcommand{\Vv}{\mathcal{V}}

\renewcommand{\vec}[1]{{\bf #1}}

\newcommand{\fun}{\textsf{fun}}
\newcommand{\rel}{\textsf{rel}}

\newcommand{\varfamily}{\Vv}
\newcommand{\funvarfamily}{\Vv^{\fun}}
\newcommand{\relvarfamily}{\Vv^{\rel}}

\newcommand{\uct}{\textsf{UCT}}
\newcommand{\proj}[2]{#1|_{#2}}

\newcommand{\strict}{pure}

\newcommand{\bool}{\textsf{bool}}

\newcommand{\step}[2]{{#2}^{\textsf{Step-#1}}}

\newcommand{\lt}{\texttt{\small LTZero}}
\newcommand{\eq}{\texttt{\small EQZero}}
\newcommand{\gt}{\texttt{\small GTZero}}
\newcommand{\add}{\texttt{\small ADD}}
\newcommand{\ite}{\texttt{\small ITE}}
\newcommand{\inp}{\texttt{\small INPUT}}

\newcommand{\lbl}{\textsf{label}}
\newcommand{\val}{\textsf{val}}
\newcommand{\lft}{\textsf{Left}}
\newcommand{\md}{\textsf{Mid}}
\newcommand{\rgt}{\textsf{Right}}

\newcommand\VRule[1][\arrayrulewidth]{\vrule width #1}

\title{A Decidable Fragment of Second Order Logic With Applications to Synthesis}

\author{P. Madhusudan}{University of Illinois, Urbana Champaign, Urbana, IL, USA}{madhu@illinois.edu}{}{}
\author{Umang Mathur}{University of Illinois, Urbana Champaign, Urbana, IL, USA}{umathur3@illinois.edu}{ https://orcid.org/0000-0002-7610-0660
}{}%{Supported by NSF CSR 1422798.}
\author{Shambwaditya Saha}{University of Illinois, Urbana Champaign, Urbana, IL, USA}{ssaha6@illinois.edu}{}{}
\author{Mahesh Viswanathan}{University of Illinois, Urbana Champaign, Urbana, IL, USA}{vmahesh@illinois.edu}{}{}%{Supported by NSF CPS 1329991.}

\authorrunning{P. Madhusudan, U. Mathur, S. Saha and M. Viswanathan}%mandatory. First: Use abbreviated first/middle names. Second (only in severe cases): Use first author plus 'et al.'

\Copyright{P. Madhusudan, Umang Mathur, Shambwaditya Saha and Mahesh Viswanathan}%mandatory, please use full first names. LIPIcs license is "CC-BY";  http://creativecommons.org/licenses/by/3.0/

\subjclass{}% mandatory: Please choose ACM 2012 classifications from https://www.acm.org/publications/class-2012 or https://dl.acm.org/ccs/ccs_flat.cfm . E.g., cite as "General and reference $\rightarrow$ General literature" or \ccsdesc[100]{General and reference~General literature}. 


\funding{This work has been supported by NSF grants 1422798, 1329991, 1138994 and 1527395.}
\EventEditors{Dan Ghica and Achim Jung}
\EventNoEds{2}
\EventLongTitle{27th EACSL Annual Conference on Computer Science Logic (CSL 2018)}
\EventShortTitle{CSL 2018}
\EventAcronym{CSL}
\EventYear{2018}
\EventDate{September 4--7, 2018}
\EventLocation{Birmingham, GB}
\EventLogo{}
\SeriesVolume{119}
\ArticleNo{30} % “LIPIcs #” (=<article-no>) goes here!
\begin{document}

\maketitle

\begin{abstract}
%!TEX root = main.tex
We propose a fragment of many-sorted second order logic called \logic
and show that checking satisfiability of
sentences in this fragment is decidable. 
\logic formulae have an
$\exists^*\forall^*$ quantifier prefix (over variables, functions and relations) 
making \logic conducive for
modeling synthesis problems. 
Moreover, \logic allows reasoning using
a combination of background theories provided that they 
have a decidable satisfiability problem for the $\exists^*\forall^*$ FO-fragment (e.g., linear
arithmetic). 
Our decision procedure reduces the
satisfiability of \logic formulae to satisfiability queries
of $\exists^*\forall^*$ formulae of each individual background theory, 
allowing us to use existing efficient SMT solvers supporting $\exists^*\forall^*$ reasoning
for these theories; hence our procedure can be seen as \emph{effectively quantified
	SMT} (\logic) reasoning.
    
\end{abstract}

\section{Introduction}
\label{sec:intro}
%!TEX root = main.tex

The goal of program synthesis is to automatically construct a program
that satisfies a given specification. This problem has received a lot
of attention from the research community in recent years~\cite{solar2006combinatorial,sygus,gulwanidimensions10}. Several different
approaches have been proposed to address this challenge (see \cite{sygus,jha2010oracle} for some
of these). One approach
to program synthesis is to reduce the problem to the satisfiability
problem in a decidable logic by constructing a sentence whose
existentially quantified variables identify the program to be
synthesized, and the inner formula expresses the requirements that the
program needs to meet.

This paper furthers this research program by identifying a
decidable \emph{second-order} logic that is suitable for encoding
problems in program synthesis. To get useful results, one needs to
constrain the semantics of functions and relations used in encoding
the synthesis problem. Therefore our logic has a set
of \emph{background theories}, where each of the background theories
is assumed to be independently axiomatized and equipped with a
solver. Finally, to leverage the advances made by logic solvers, our aim
is to develop a decision procedure for our logic that
makes \emph{black-box} calls to the decision procedures 
(for $\exists^*\forall^*$ satisfiability) for the
background theories.

With the above goal in mind, let us describe our logic. It is
a \emph{many-sorted} logic that can be roughly described as
an \emph{uninterpreted combination of theories} (UCT)~\cite{natproof17}.
A UCT has a many-sorted universe where there is a special sort
$\sigma_0$ that is declared to be a \emph{foreground} sort, while the other
sorts ($\sigma_1, \ldots \sigma_n$) are declared to be background sorts.
We assume that there is some fixed signature of functions, relations,
and constants over each individual background sort that is purely over
that sort. Furthermore, we assume that each background sort $\sigma_i$
($i>0$) comes with an associated \emph{background theory} $T_i$; $T_i$
can be arbitrary, even infinite, but is constrained to formulae
with functions, relations and constants that only involve the
background sort $\sigma_i$.
Our main contribution is a decidability result for the satisfiability
problem modulo these %background 
theories for boolean combinations of
sentences of the form
\begin{align}
% \vspace{-0.5in}
\label{eqn:efblock}
(\exists \vec{x}) (\exists \mathbfcal{R}) (\exists \vec{F}) 
(\forall \vec{y}) (\forall \mathbfcal{P}) (\forall \vec{G}) \psi, %\quad \text{where}
\end{align}
%
% \vspace{-0.5cm}
% where,
% $\vec{x}$ is a set of existentially quantified first order 
% variables that can admit values in any of the sorts
% (background or foreground);
% $\mathbfcal{R}$ is a set of existentially quantified relation
% variables, whose arguments are restricted to be over the foreground
% sort $\sigma_0$;
% $\vec{F}$ is a set of existentially quantified function variables, 
% which take as arguments elements from the foreground sort
% $\sigma_0$, and return a value in any of the background sorts
% $\sigma_i$;
% $\vec{y}$ is a set of universally quantified first order variables 
% over any of the sorts;
% $\mathbfcal{P}$ is a set of universally quantified relation 
% variables, whose arguments could be of any of the sorts; and 
% $\vec{G}$ is a set of universally quantified function variables, 
% whose arguments can be from any sort and could return values of any sort.
% where 
\begin{itemize}
\item $\vec{x}$ is a set of existentially quantified first order 
  variables. These variables can admit values in any of the sorts
  (background or foreground);
\item $\mathbfcal{R}$ is a set of existentially quantified relational 
  variables, whose arguments are restricted to be over the foreground
  sort $\sigma_0$;
\item $\vec{F}$ is a set of existentially quantified function variables, 
  which take as arguments elements from the foreground sort
  $\sigma_0$, and return a value in any of the background sorts
  $\sigma_i$;
\item $\vec{y}$ is a set of universally quantified first order variables 
  over any of the sorts;
\item $\mathbfcal{P}$ is a set of universally quantified relational 
  variables, whose arguments could be of any of the sorts; and 
\item $\vec{G}$ is a set of universally quantified function variables, 
  whose arguments can be from any sort and could return values of any sort.
\end{itemize}
Thus our logic has sentences with prefix $\exists^*\forall^*$,
allowing for quantification over both first order variables and
second-order variables (relational and functional). To obtain
decidability, we have carefully restricted the sorts (or \emph{types}) of second-order
variables that are existentially and universally quantified, as described above.

Our decidability result proceeds as follows. By crucially exploiting
the disjointness of the universes of background theories and through a
series of 
% simple observations and 
transformations,
% like
% Ackermanization, decision procedures similar to the EPR logic (or the
% Bernays-Sch\"onfinkel-Ramsey class) that exploit the finite model
% property, 
we reduce the satisfiability problem for our logic to the
satisfiability of several pure $\exists^*\forall^*$ \emph{first-order} logic
formulas over the individual background theories $T_1,\ldots T_n$.
Consequently, if the background theories admit (individually) a
decidable satisfiability problem for the first-order $\exists^*\forall^*$
fragment, then satisfiability for our logic
is \emph{decidable}. Examples of such background theories include Presburger
arithmetic, the theory of real-closed fields, and the theory of linear
real arithmetic. Our algorithm for satisfiability
% , therefore, 
makes finitely many black-box calls to the engines for the
individual background theories.

\subsection*{Salient aspects of our logic and our decidability result}

\noindent{\bf Design for decidability:} 
Our logic is defined to carefully avoid the undecidability that
looms in any logic of such power. 
We do not know of any decidable second-order
logic fragment that supports background theories such as arithmetic and
uninterpreted functions. 
While \emph{quantifier-free} decidable logics can be combined to get
decidable logics using Nelson-Oppen combinations~\cite{no79},
or local theory extensions~\cite{localtheory2013},
combining quantified
logics is notoriously hard, 
and there are only few restricted classes of first-order
logic that are known to be decidable.
% For example, stratified fragments as in~\cite{abadi2010decidable,Padon:2017:PME:3152284.3140568},
% restrict the communication between existentially and universally quantified variables,
% and also rely on finite model property of the quantifier free formulae
% of the component pure theories.
% \ucomment{
% Often, the decision procedures for such combinations rely on 
% reduction to satisfiability of the quantifier free fragment of the individual theories~\cite{abadi2010decidable,}}

% \noindent{\bf Design for decidability:} 
% Our logic is also defined to carefully avoid the undecidability that
% looms in any such logic of such power. First, note that
% while \emph{quantifier-free} decidable logics can be combined to get
% decidable logics using Nelson-Oppen combinations~\cite{no79}, combining quantified
% logics are notoriously hard. 
% Our design choice
% forces \emph{communication} between theories using the foreground
% sort, and further keeping the universes of the different
% sorts \emph{disjoint} allows a decidable combination of
% $\exists^* \forall^*$ theories.

Our design choice forces \emph{communication} between theories using the foreground
sort, keeping the universes of the different
sorts \emph{disjoint}, which allows a decidable combination of
$\exists^* \forall^*$ theories.
We emphasize that, unlike existing work on quantified first-order theories that
are decidable by reduction to quantifier-free SMT, our logic allows existential
and universal quantification over the background theories as well, and the
decision procedure reduces satisfiability to $\exists^* \forall^*$ fragment of
the underlying theories. Our result can hence be seen as a decidable combination
of $\exists^* \forall^*$ theories that further supports second-order quantification.
% ,
% while Nelson-Oppen combinations combine quantifier-free theories.

\medskip

\noindent{\bf Undecidable Extensions:} 
We show that our logic is on the edge of the decidability barrier, by showing that
lifting some of the restrictions we have will render the logic undecidable.
In particular, we show that if we allow  outer existential quantification over functions
(which is related to the condition demanding that all function variables are universally
quantified in the inner block of quantifiers),  then satisfiability of the logic is undecidable.
Second, if we lift the restriction that the underlying background sorts are pairwise disjoint,
then again the logic becomes undecidable. 
The design choices that we have made hence seem
crucial for decidability.

\medskip

\noindent{\bf Expressing Synthesis Problems:} 
Apart from decidability, a primary motivational design principle of our logic is to
express \emph{synthesis} problems. Synthesis problems
typically can be expressed in $\exists^* \forall^*$ fragments, where
we ask whether there exists an object of the kind we wish to synthesize
(using the block of existential quantifiers) such that the object
satisfies certain properties (expressed by a universally quantified
formula). For instance, if we are synthesizing a program snippet that
is required to satisfy a Hoare triple (pre/post condition), we can encode this by
asking whether there is a program snippet such that for all values of
variables (modeling the input to the snippet), the verification
condition corresponding to the Hoare triple holds.  In this context,
the existentially quantified variables (first order and second order)
can be used to model program snippets. Furthermore, since our logic allows
second-order universal quantification over \emph{functions}, we can
model aspects of the program state that require \emph{uninterpreted
functions}, in particular pointer fields that model the heap. 
%As we
%will illustrate in this paper, we can use our framework to synthesize
%even \emph{recursive programs} given pre/post conditions for them and
%for properties involving heaps and recursive definitions.

\medskip

\noindent{\bf Evaluation on Synthesis Problems:} 
We illustrate the applicability of our logic for two classes of
synthesis problems.
The first class involves
synthesizing \emph{recursive programs} that work over inductive
data-structures.  Given the precise pre/post condition for the program
to be synthesized, we show how to model recursive program synthesis by
synthesizing only a straight-line program (by having the output of
recursive calls provided as inputs to the straightline program). 
The verification condition of the program requires universal quantification
over both scalar variables as well as heap pointers, modeled as uninterpreted functions.
Since such verification-conditions are already very expressive (even for the purpose of verification), 
we adapt a technique in the literature called \emph{natural proofs}~\cite{natproof17,qiu2013natural,pek2014natural},
that soundly abstracts the verification condition to a decidable theory.
This formulation still has universal quantification over variables and functions,
and combines standard background theories such as arithmetic and theory of uninterpreted functions. 
We then show
that synthesis of bounded-sized programs (possibly involving 
integer constants that can be unbounded) can be modeled in our logic. In this
modeling, the universal quantification over functions plays a crucial
role in modeling the pointers in heaps, and modeling uninterpreted
relations that capture inductive data-structure predicates (such as
{\tt lseg}, {\tt bstree}, etc.).
%
% We illustrate the above class of examples by showing the
% synthesis of a recursive program that searches for a key in a singly
% linked list. 
%Though we do not have an efficient implementation of a
%decision procedure for our logic, we use an existing \emph{enumeraion
%based counter-example guided synthesis engine} to solve the synthesis
%problem and illustrate our decision procedure.

The second class of synthesis involves taking a recursive definition
of a function, and synthesizing a non-recursive (and iteration free)
function equivalent to it. In our modeling, the existential
quantification over the foreground sort as well as the background sort
of integers is utilized, as the synthesized function involves
integers. 

The crux of our contribution, therefore, is providing a decidable logic 
that can express synthesis problems succinctly. 
% We believe such a logic
% can be useful for researchers working on practical 
% applications of program/expression synthesis. Implementing the
% decision procedure effectively can be then engineered efficiently by
% researchers working on logic engines. 
%
Such a logic promises to provide a useful interface between researchers
working on practical synthesis applications and researchers
working on engineering efficient tools for solving them,
similar to the role SMT plays in verification.
\vspace{-0.1in}

\section{Motivating \logic for synthesis applications}
\label{sec:example}
%!TEX root = main.tex
% \DeclareMathAlphabet\mathbfcal{OMS}{cmsy}{b}{n}
% Program synthesis problems can be conveniently encoded in \logicWithComma
% which we define formally in Section~\ref{sec:logic}.
% \logic allows existential and universal quantification
% over variables, relations and functions over many-sorted signatures
% (with some restrictions on the \emph{types} of existentially
% quantified relations and functions, in order to ensure
% decidability). Sentences in \logic are of the form
% \begin{align}
% \label{eqn:efblock}
% \phi = \underbrace{(\exists \vec{x}) (\exists \mathbfcal{R}) (\exists \vec{F})}_{\exists\text{-Block}} \, (\underbrace{\forall \vec{y}) (\forall \mathbfcal{P}) (\forall \vec{G})}_{\forall\text{-Block}} \, \psi
% \end{align}
% %
% There is an initial block of existential quantifiers
% ($\exists$-Block), followed by universal quantifiers
% ($\forall$-Block).  In Section~\ref{sec:dec}, we show that if the
% constituent background theories admit decision procedures for
% satisfiability of their respective $\exists^*\forall^*$ fragments,
% then so does the fragment \logicWithPeriod  Examples of such background
% theories include Presburger arithmetic, linear arithmetic of
% rationals, and the theory of real closed fields.
% % \ucomment{what else?}, etc.,.
% \vspace{-0.2cm}
In program synthesis, the goal is to search for programs, typically of bounded
size, that satisfy a given specification.  The $\exists$-Block of an \logic formula
can be used to express the search for the syntactic program.  
The inner formula, then, must interpret the semantics of this syntactic program,
and express that it satisfies the specification.
If the specification is a universally
quantified formula, then, we can encode the synthesis problem in \logicWithPeriod

One of the salient features of the fragment \logic is the
ability to universally quantify over functions and relations.
Often, specifications for programs, such as those
that manipulate heaps, involve a universal quantification over 
uninterpreted functions (that model pointers).
\logic aptly provides this functionality, while still
remaining within the boundaries of decidability.
Further, \logic supports combination of background theories/sorts;
existential quantification over these sorts can thus be used
to search for programs with arbitrary elements from these background sorts.
As a result, the class of target programs that can be expressed by 
an \logic formula is infinite.
Consequently, when our decision procedure returns unsatisfiable,
we are assured that no program (from an infinite class of programs) exists, 
(most CEGIS solvers for program synthesis cannot provide such a guarantee.)

% Let us now throw some light on the many-sorted aspect of \logicWithPeriod  This
% is aspect is particularly useful in synthesis of programs that
% manipulate objects with a wide variety of \emph{types} (such as
% integers, arrays, lists, strings, sets, heaps, etc.,).  Our
% logic \logic allows existential quantification over variables of many
% different sorts, thus allowing for synthesis of a richer class of
% programs.
% We believe this and other unique aspects of \logic will
% be useful in modeling synthesis of more general models
% of computation such as reactive and hybrid systems.

% We demonstrate the applicabil
We now proceed to give a concrete example of a synthesis problem
which will demonstrate the power of \logicWithPeriod
Consider the specification of the following function $M_{\textsf{three}}$,
which is a slight variant of the classical McCarthy's $91$ function~\cite{manna1969properties}, 
whose specification is given below.
%
% \vspace{-0.8cm}
%
\begin{align}
\label{eqn:mthree}
M_\textsf{three}(n) = 
\begin{cases}
n - 30 & \text{ if } n > 13 \\
M_\textsf{three}(M_\textsf{three}(M_\textsf{three}(n+61))) & \text{ otherwise }
\end{cases}
\end{align}
% \begin{wrapfigure}{l}{0.55\textwidth}
% \vspace{-0.35in}
% \scriptsize
% \begin{empheq}[box=\fbox]{flalign*}
% \mathtt{term} := & \;\;  \mathtt{c} \;\;  | \;\;  \mathtt{( x + c)} \;\; | \;\; \mathtt{ite( pred ,\, term,\, term)} \\
% \mathtt{pred} := & \;\; \mathtt{( x < c)} \;\; | \;\; \mathtt{( x = c)} \;\; | \;\; \mathtt{( x > c )}
% \end{empheq}
% % \begin{align*}
% % \mathtt{term} := & \;\;  \mathtt{c} \;\;  | \;\;  \mathtt{( x + c)} \;\; | \;\; \mathtt{ite( pred ,\, term,\, term)} \\
% % \mathtt{pred} := & \;\; \mathtt{( x < c)} \;\; | \;\; \mathtt{( x = c)} \;\; | \;\; \mathtt{( x > c )}
% % \end{align*}
% \vspace{-0.2in}
% \caption{Grammar for $M_{\textsf{three}}$.
% % $\mathtt{ite}(\cdot, \cdot, \cdot)$ stands for \texttt{if-then-else}.
% }
% \vspace{-0.1in}
% \label{fig:gram}
% \end{wrapfigure}
% \input{parse_tree}
We are interested in synthesizing a straight line program
that implements the recursive function $M_\textsf{three}$, and
can be expressed as a \texttt{term} over the grammar specified in  Figure~\ref{fig:gram}.

% \vspace{-0.4cm}

%!TEX root = main.tex

% \begin{figure}
% \centering
% \begin{minipage}{.55\textwidth}
%   \scriptsize
% \begin{empheq}[box=\fbox]{flalign*}
% \mathtt{term} := & \;\;  \mathtt{c} \;\;  | \;\;  \mathtt{( x + c)} \;\; | \;\; \mathtt{ite( pred ,\, term,\, term)} \\
% \mathtt{pred} := & \;\; \mathtt{( \mathtt{term} < 0 )} \;\; | \;\; \mathtt{( \mathtt{term} = 0)} \;\; | \;\; \mathtt{( \mathtt{term} > 0 )}
% \end{empheq}
%   \captionof{figure}{Grammar for $M_{\textsf{three}}$.
% $\mathtt{ite}(\cdot, \cdot, \cdot)$ stands for \texttt{if-then-else}.}
%   \label{fig:gram}
% \end{minipage}
% \hfill
% \begin{minipage}{.35\textwidth}
%   \scalebox{0.7}{
% \begin{tikzpicture}[sibling distance=5em,
%   every node/.style = {shape=rectangle, rounded corners,
%     fill=black!10, draw=black, align=center}]]
%   \node {$n_0$}
%     child { 
%       node {$n_{00}$}
%       child { node {$n_{000}$} } 
%       child { node {$n_{001}$} } 
%       child { node {$n_{002}$} } 
%     }
%     child { 
%       node[draw=none, fill=none] {$\vdots$}
%     }
%     child { 
%      node[draw=none, fill=none] {$\vdots$}
%     };
% \end{tikzpicture}
% }
% \label{fig:skeleton}
% \captionof{figure}{Program skeleton}
% \end{minipage}%
% \end{figure}
% \vspace{-0.3cm}
\begin{figure}[t]
\centering
\begin{subfigure}{.55\textwidth}
  \scriptsize
\begin{empheq}[box=\fbox]{flalign*}
\mathtt{term} := & \;\;  \mathtt{c} \;\;  | \;\;  \mathtt{( x + c)} \;\; | \;\; \mathtt{ite( pred ,\, term,\, term)} \\
\mathtt{pred} := & \;\; \mathtt{( \mathtt{term} < 0 )} \;\; | \;\; \mathtt{( \mathtt{term} = 0)} \;\; | \;\; \mathtt{( \mathtt{term} > 0 )}
\end{empheq}
  \caption{
  \label{fig:gram}
  Grammar for $M_{\textsf{three}}$.
$\mathtt{ite}(\cdot, \cdot, \cdot)$ stands for \texttt{if-then-else}.}
\end{subfigure}
\hfill
\begin{subfigure}{.35\textwidth}
  \scalebox{0.7}{
\begin{tikzpicture}[sibling distance=5em,
  every node/.style = {shape=rectangle, rounded corners,
    fill=black!10, draw=black, align=center}]]
  \node {$n_0$}
    child { 
      node {$n_{00}$}
      child { node {$n_{000}$} } 
      child { node {$n_{001}$} } 
      child { node {$n_{002}$} } 
    }
    child { 
      node[draw=none, fill=none] {$\vdots$}
    }
    child { 
     node[draw=none, fill=none] {$\vdots$}
    };
\end{tikzpicture}
}
\caption{Program skeleton}
\label{fig:skeleton}
\end{subfigure}%
\caption{
\label{fig:enc_mthree}
Synthesizing $M_\textsf{three}$ using \logicWithPeriod}
\end{figure}
% \vspace{-0.3cm}

% \begin{align*}
% \mathtt{term} := & \;\;  \mathtt{c} \;\;  | \;\;  \mathtt{( x + c)} \;\; | \;\; \mathtt{ite( predicate ,\, term,\, term)} \\
% \mathtt{predicate} := & \;\; \mathtt{( x < c)} \;\; | \;\; \mathtt{( x = c)} \;\; | \;\; \mathtt{( x > c )}
% \end{align*}
% \ucomment{Maybe say that since terms obtained using these are the same when you also have boolean combination of predicates; so the above grammar is more general than it looks like}.

% Let us see how to encode this synthesis problem in \logicWithPeriod
Here, we only briefly discuss how to encode this synthesis problem in \logicWithComma
and the complete details can be found in Appendix~\ref{app:mthree}.
First, let us fix the maximum height of the term we are looking for,
say to be 2.
Then, the program we want to synthesize can be represented 
as a tree of height at most $2$ such that every node in 
the tree can have $\leq 3$ child nodes 
(because the maximum arity of any function 
in the above grammar is $3$, corresponding to $\texttt{ite}$).
The skeleton of such an expression tree is shown in Figure~\ref{fig:skeleton}.
Every node in the tree is named according to its path from the root node.

The synthesis problem can then be encoded as the following formula
\vspace{-0.2cm}
\begin{align*}
\phi_{M_\textsf{three}} \equiv \, & (\exists n_0, n_{00}, n_{01}, \ldots n_{022} : \sigma_0) \; (\exists \lft, \md, \rgt : \sigma_0 \sigma_0) \\
& (\exists \add,\ite,\lt,\eq,\gt,\inp, C_1, C_2, C_3 : \sigma_\lbl)\; \\
& (\exists {c_1, c_2, c_3 : \mathbb{N}}) \; (\exists f_\lbl : \sigma_0, \sigma_\lbl) \; \\
& \quad\quad  \varphi_\textsf{well-formed} \land (\forall x : \mathbb{N}) (\forall g_\val : \sigma_0, \mathbb{N}) \;  (\varphi_{\textsf{semantics}} \implies \varphi_{\textsf{spec}})
\end{align*}
Here, the nodes $n_0, n_{00}, \ldots$ are elements of the foreground sort $\sigma_0$.
The binary relations $\lft, \md, \rgt$ over the foreground sort
will be  used to assert that
a node $n$ is the left,middle, right child respectively of node $n'$ :
$\lft(n', n)$, $\md(n', n)$, $\rgt(n', n)$.
The operators or \emph{labels} for nodes belong to the background sort $\sigma_\lbl$,
and can be one of $\add \, (+)$, $\ite \, (\mathtt{ite})$, $\lt \, (< 0)$,
$\gt \, (> 0)$, $(\eq \, (= 0))$, $\inp$ (denoting the input to our program),
or constants $C_1, C_2, C_3$ (for which we will synthesize natural constants $c_1, c_2, c_3$
in the (infinite) background sort $\mathbb{N}$).
The function $f_\lbl$ assigns a label to every node in the program,
and the formula $\varphi_\textsf{well-formed}$ asserts some sanity conditions:
\vspace{-0.2cm}
\begin{align*}
\varphi_\textsf{well-formed}  \equiv & \bigwedge\limits_{\rho \neq \rho'} n_\rho \neq n_{\rho'} \; 
\land \lft(n_0, n_{00}) \land \bigwedge\limits_{\rho \neq 00} \neg(\lft(n_0, n_\rho))) \land \cdots\\
& \land \neg (\add = \ite) \land \neg (\add = \lt) \land \cdots \land \neg(C_1 = C_3) \land \neg (C_2 = C_3) \\
& \land \bigwedge\limits_{\rho}   (f_\lbl(n_\rho) {=} \add) \lor (f_\lbl(n_\rho) {=} \ite) \lor \cdots \lor (f_\lbl(n_\rho) {=} C_3 )
\end{align*}
% \vspace{-0.2cm}
The formula $\varphi_\textsf{semantics}$ asserts that the `meaning'
of the program can be inferred from the meaning of the components of the program.
We will use the function $g_\val$, that assigns value to nodes from $\mathbb{N}$,
for this purpose :
\begin{align*}
\varphi_\textsf{semantics} \equiv \bigwedge\limits_{\rho, \rho_1, \rho_2}
\begin{bmatrix}
% \begin{array}{rcl}
\Big(f_\lbl(n_\rho) = \add \land \lft(n_\rho, n_{\rho_1})  \land \md(n_\rho, n_{\rho_2}) \Big) \\
\implies g_\val(n_\rho) = g_\val(n_{\rho_1}) + g_\val(n_{\rho_2}) \Big) \\
\vdots \\
\land \; f_\lbl(n_\rho) = C_3  \implies g_\val(n_\rho) = c_3
\end{bmatrix}
\end{align*}
% \vspace{-0.2cm}

Finally, the formula $\varphi_\textsf{spec}$ expresses the specification of the program
as in Equation~\eqref{eqn:mthree}.
A complete description is provided in Appendix~\ref{app:mthree}.

Observe that the formula $\phi_{M_\textsf{three}}$ has
existential and universal quantification over functions and relations,
as allowed by our decidable fragment \logicWithPeriod 
The existentially quantified functions map the foreground sort
$\sigma_0$ to one of the background sorts, and the 
existentially quantified relations span only over the foreground sort.

% For the purpose of evaluating the effectiveness of our logic,
We encoded the above \logic formula in the 
\texttt{z3}~\cite{DeMoura:2008:ZES:1792734.1792766} SMT solver (see Section~\ref{sec:exp} for details), 
which synthesized the expression $\mathtt{fun(n)} = \mathtt{ite(n > 13, n-30, -16)}$.
 In Section~\ref{sec:exp}, we show that we can synthesize
a large class of such programs amongst others.%`$\mathtt{fun(n)} = \mathtt{ite(n > 13, n-30, -16)}$'.
% \vspace{-0.2cm}
% \begin{align*}
% \mathtt{fun(n)} = \mathtt{ite(n > 13, n-30, -16)}
% % \vspace{-0.1cm}
% \end{align*}
% Indeed this function satisfies the specification
% of the function $M_\textsf{three}$, thus implying
% the effectiveness of our approach.

\section{Many-sorted Second Order Logic and the \logic Fragment}

\label{sec:logic}
%!TEX root = main.tex

We briefly recall the syntax and semantics of general many-sorted second order logic, and then
present the $\logic$~fragment of second order logic.

% \vspace{-0.4cm}

\subsection*{Many-sorted second-order logic}
A many-sorted signature is a tuple
$
\Sigma = (\Ss, \Ff, \Rf, \varfamily, \funvarfamily, \relvarfamily)
$
where, $\Ss $ is a nonempty finite set of sorts, 
$\Ff$, $\Rf$, $\varfamily$, $\funvarfamily$, $\relvarfamily$
are, respectively, sets of function symbols,
relation symbols, first order variables,
function variables and relation variables.
Each variable $x \in \varfamily$ is associated with a sort $\sigma \in \Ss$, 
represented as $x\!:\!\sigma$.
Each function symbol or function variable also has an associated type
$(w,\sigma) \in \Ss^*\times \Ss$,
and each relation symbol and relation variable has a type $w \in \Ss^+$.
% $\varfamily$ is a set of (first order) variables, each with an associated sort $\sigma \in \Ss$,
% $\Ff$ is a set of function symbols, each associated with a type $w\in \Ss^*\times\Ss$,
% and  $\Rf$ is a set of relation symbols, each also associated with a type 
% $w\in \Ss^+$.
% $\funvarfamily$ is a set of second-order \emph{function variables},
% each with an associated type $w\in \Ss^*\times\Ss$, and 
% $\relvarfamily$ is a set of second-order \emph{relational variables}
% with types $w\in \Ss^+$.
We assume that the set of symbols in $\Ff$ and $\Rf$ are either finite
or countably infinite, and that $\varfamily$, $\funvarfamily$, and
$\relvarfamily$ are all countably infinite.  Constants are modeled
using $0$-ary functions.
We say that $\Sigma$ is \emph{unsorted} if $\Ss$ consists of a single sort.

% \begin{itemize}
% 	\item $\Ss $ is a finite set of sorts
% 	\item $\Ff = \{F_w\}_{w\in \Ss^*\times\Ss}$ is a $\Ss^*{\times}\Ss$ indexed family of function symbols
% 	\item $\Rf = \{\Rr_w\}_{w \in \Ss^+}f$ is a $\Ss^+$ indexed family of relation symbols
% 	\item $\varfamily = \{V_s\}_{s\in \Ss}$ is a family of (first order) variables
% 	\item $\funvarfamily = \{V^\fun_w\}_{w\in \Ss^*\times\Ss}$ is a family of functional variables
% 	\item $\relvarfamily = \{V^\rel_w\}_{w\in \Ss^*\times\Ss}$ is a family of relational variables
% \end{itemize}
%\ucomment{Variable sets is (countably) infinite}.
%\ucomment{Also say that constants are just 0-ary functions}.
%\ucomment{$\Sigma$ is single sorted if $\Ss$ is singleton.}
%\ucomment{Say that we assume the non-degenerate case $|\Ss| > 0$.}

Terms over a many-sorted signature $\Sigma$ have an associated sort
and are inductively defined by the grammar
\vspace{-0.3cm}
\[
t\!:\!\sigma \quad :=  \quad x\!:\!\sigma \quad |\quad f(t_1\!:\!\sigma_1, t_2\!:\!\sigma_2, \ldots, t_m\!:\!\sigma_m) \quad |\quad F(t_1\!:\!\sigma_1, t_2\!:\!\sigma_2, \ldots, t_n\!:\!\sigma_n)
\]
% \vspace{-0.1cm}
\noindent
where $f\!:\!(\sigma_1\sigma_2\cdots\sigma_m,\sigma) \in \Ff$, and
$F\!:\!(\sigma_1\sigma_2\cdots\sigma_n,\sigma) \in \funvarfamily$.
% \vspace{-0.2cm}
% 
Formulae over $\Sigma$ are inductively defined as
% \vspace{-0.5cm}
\begin{align*}
\phi \quad := \quad &\bot \quad |\quad \phi \Rightarrow \phi \quad |\quad t\!:\!\sigma = t'\!:\!\sigma \quad|\quad R(t_1\!:\!\sigma_1, t_2\!:\!\sigma_2, \ldots, t_m\!:\!\sigma_m) \quad |\quad \\
&  \Rr(t_1\!:\!\sigma_1, t_2\!:\!\sigma_2, \ldots, t_n\!:\!\sigma_n) \quad | \quad (\exists x\!:\!\sigma) \, \phi \quad |\quad (\exists F\!:\! w,\sigma) \, \phi \quad| \quad (\exists \Rr'\!:\! w) \,  \phi
\end{align*}

% \vspace{-0.1cm}

\noindent
where $R\!:\!(\sigma_1\sigma_2\cdots\sigma_m) \in \Rf$, $\Rr, \Rr'$
are relation variables, $F$ is a function variable, of appropriate
types.
%
% , $\Rr\!:\!(\sigma_1\sigma_2\cdots\sigma_n) \in \relvarfamily$.
% $\Rr'\!:\!w \in \relvarfamily$, and $F\!:\!(w, \sigma) \in \funvarfamily$.%, $y$ is free in $\phi_3$ and $x$ is free in $\phi_4$ .
%
Note that equality is allowed only for terms of same sort.
%
%\ucomment{Define free variable, sentence, FO restriction.}
%
A formula is said to be \emph{first-order} if it does not use any
function or relation variables.

The semantics of many sorted logics are described using many-sorted
structures.  A $\Sigma$-structure is a tuple $
\Mm = (\Uu, \Ii)
$ where $\Uu = \{M_\sigma\}_{\sigma\in \Ss}$ is a collection of
pairwise disjoint $\Ss$ indexed universes, and $\Ii$ is an
interpretation function that maps each each variable $x:\sigma$ to an
element in the universe $M_\sigma$, each function symbol and each
function variable to a function of the appropriate type on the
underlying universe.  Similarly, relation symbols and relation
variables are also assigned relations of the appropriate type on the
underlying universe. For an interpretation $\Ii$, as is standard, we
use $\Ii[c^x/x]$ to denote the interpretation that maps $x$ to $c^x$,
and is otherwise identical to $\Ii$. For function variable $F$ and
relation variable $\Rr$, $\Ii[f^F/F]$ and $\Ii[R^\Rr/\Rr]$ are defined
analogously.

%
%and each functional variable in $V^\fun_{\sigma_1\sigma_2\ldots\sigma_k\times\sigma}$
%to an element in $[M_{\sigma_1}\times M_{\sigma_2}\ldots M_{\sigma_k} \to M_\sigma]$,
%and, each relation symbol in $\Rr_{\sigma_1\sigma_2\ldots\sigma_k}$
%and relational variable in $V^\rel_{\sigma_1\sigma_2\ldots\sigma_k}$ 
%to a subset of $M_{\sigma_1}\times M_{\sigma_2}\ldots M_{\sigma_k}$.
% \begin{itemize}
% 	\item $\Uu$ is a $\Ss$ indexed family of universes : $\Uu = \{M_\sigma\}_{\sigma\in \Ss}$
% 	\item $\Ii$ is an interpretation function such that
% 	\begin{itemize}
% 		\item for every function symbol $f \in F_{\sigma_1\sigma_2\ldots\sigma_k\times\sigma}$, we have $\Ii(f) \in [M_{\sigma_1}\times M_{\sigma_2}\ldots M_{\sigma_k} \to M_\sigma]$
% 		\item for every relation symbol $R \in \Rr_{\sigma_1\sigma_2\ldots\sigma_k}$, we have $I_s(R) \subseteq M_{\sigma_1}\times M_{\sigma_2}\ldots M_{\sigma_k}$
% 		\item for every variable $x:\sigma \in X$, we have $\Ii(x) \in M_\sigma$
% 		\item for every functional variable $y:\sigma_1\sigma_2\ldots\sigma_k\to\sigma \in Y$, we have $\Ii(y) \in [M_{\sigma_1}\times M_{\sigma_2}\ldots M_{\sigma_k} \to M_\sigma]$
% 	\end{itemize}
% \end{itemize}
%\ucomment{Interpretation extends to terms.}

Interpretation of terms in a model is the usual one obtained by
interpreting variables, functions, and function variables using their
underlying interpretation in the model; we skip the details.
The satisfaction relation $\Mm \models \phi$ is also defined in the usual sense,
and we will skip the details.

% For $\Mm = (\Uu, \Ii)$, with $\Uu = \{M_\sigma\}_{\sigma\in \Ss}$, the
% satisfaction relation $\models$ is defined as:
% \begin{center}
% {
% \begin{tabular}{lcr}
% $\Mm \models \bot$ & \quad & for no $\Mm$ \\
% $\Mm \models \phi_1 \implies \phi_2$ & iff & $\Mm \not\models \phi_1$ or $\Mm \models \phi_2$ \\
% $\Mm \models (t_1\!:\!\sigma = t_2\!:\!\sigma) $ & iff & $\Ii(t_1) = \Ii(t_2)$ \\
% $\Mm \models R(t_1\!:\!\sigma_1, t_2\!:\!\sigma_2, \ldots, t_n\!:\!\sigma_n)$ \quad & iff & $(\Ii(t_1), \Ii(t_2), \ldots, \Ii(t_n)) \in \Ii(R)$\\
% $\Mm \models \Rr(t_1\!:\!\sigma_1, t_2\!:\!\sigma_2, \ldots, t_n\!:\!\sigma_n)$ & iff & $(\Ii(t_1), \Ii(t_2), \ldots, \Ii(t_n)) \in \Ii(\Rr)$\\
% $\Mm \models \exists (x\!:\!\sigma) \, \phi$ & iff & \quad $(\Uu, \Ii[c^x/x]) \models \phi$ for some $c^x \in M_\sigma$\\
% $\Mm\models \exists (F\!:\!\sigma_1\sigma_2\ldots\sigma_n\times\sigma) \, \phi$ & iff & $(\Uu, \Ii[f^F/F]) \models \phi$ for some  \\
%  & & $f^F \in [M_{\sigma_1}\times M_{\sigma_2}\cdots M_{\sigma_n}\to M_{\sigma}]$  \\
%  $\Mm\models \exists (\Rr\!:\!\sigma_1\sigma_2\ldots\sigma_n) \, \phi$ & iff & $(\Uu, \Ii[R^\Rr/\Rr]) \models \phi$ for some  \\
%  & & $R^\Rr \subseteq M_{\sigma_1}\times M_{\sigma_2}\cdots M_{\sigma_n}$  \\
% \end{tabular}
% }
% \end{center}

A first-order theory is a tuple $T = (\Sigma_T, \Aa_T)$, where
$\Aa_T$ is a set of (possibly infinite) first-order sentences.
Theory $T$ is \emph{complete} if every sentence $\alpha$ or
its negation is entailed by $\Aa_T$, i.e., either every model
satisfying $\Aa_T$ satisfies $\alpha$, or every model satisfying
$\Aa_T$ satisfies $\neg \alpha$. A theory $\Aa_T$ is \emph{consistent}
if it is not the case that there is a sentence $\alpha$ such that both
$\alpha$ and $\neg\alpha$ are entailed.

\subsection*{The logic \logic}

We now describe \logicWithComma the fragment of many-sorted second order logic
that we prove decidable in this paper and that we show can model
synthesis problems.

%\ucomment{Define projection to string of sorts}.

Let $\Sigma = (\Ss, \Ff, \Rf, \varfamily, \funvarfamily, \relvarfamily)$
be a many sorted signature.
$\Sigma$ is a \emph{{\strict} signature} if (a) the type of every function symbol
and every relation symbol is over a single sort (however, function variables
and relation variables are allowed to mix sorts), (b) there is a special sort
$\sigma_0$ (which we call the \emph{foreground sort}, while other sorts $\sigma_1, \ldots, \sigma_k$
are called \emph{background sorts}) and (c) there are no function or relation
symbols involving $\sigma_0$.

%$\sigma \in \Ss$ if for all $w \in \sigma^*$, we have 
%$\proj{\Ff}{w} = \proj{\Rf}{w} = \proj{\funvarfamily}{w} = \emptyset$.
%$\Sigma$ is \ucomment{\fo}~with respect to sort $\sigma\in \Ss$
%if for every $w \in \sigma^*$, we have
%$\proj{\funvarfamily}{w} = \proj{\relvarfamily}{w} = \emptyset$.

%An combin (UCT) is defined over a signature
%that is \strict, has a special sort $\sigma_0$ which is \mt (called the foreground
%sort; the other sorts $\sigma_1, \ldots \sigma_k$) are called \emph{background sorts}), and 
%
%
%$\uct$ if it is \strict, and
%there is a special sort $\sigma_0 \in \Ss$ such that 
%$\Sigma$ is \mt~with respect to $\sigma_0$.
%We call $\sigma_0$ as the \emph{foreground} sort, and all other
%sorts, the \emph{background} sorts of $\Sigma$.
%For a $\uct$ signature $\Sigma$, its boolean completion
%$\hat{\Sigma}$ is also a $\uct$ signature, with the same foreground sort,
%and an additional background sort $\bool$.

The fragment \logic is the set of sentences defined
over a {\strict} signature $\Sigma$, with foreground sort $\sigma_0$ and background sorts $\sigma_1, \ldots \sigma_k$,
by the following grammar
\vspace{-0.2cm}
\[
\phi \quad := \quad \varphi \quad |\quad \exists (x:\sigma) \, \phi \quad |\quad (\exists \Rr:w) \, \phi \quad |\quad (\exists F:w,\sigma_i) \, \phi
\]
where, $\sigma \in \Ss$, $w \in \sigma_0^+$ (i.e., only foreground sort), $1\leq i\leq k$, and
$\varphi$ is a universally quantified formula defined by the grammar
\[
\varphi \quad :=\quad \psi \quad |\quad \forall (y:\sigma) \, \varphi \quad |\quad (\forall \Rr:w')\, \varphi \quad |\quad (\forall F:w',\sigma) \, \varphi \]
where, $\sigma \in \Ss$, $w' \in \Ss^+$, and $\psi$ 
is quantifier free over $\Sigma$.

The formulas above consist of an existential quantification block
followed by a universal quantification block. 
The existential block can have first-order variables of any sort, relation variables that
are over the foreground sort only, and function variables that map
tuples of the foreground sort to a background sort. The inner
universal block allows all forms of quantification --- first-order
variables, function variables, and relation variables of all possible
types. The inner formula is quantifier-free. We will retrict our
attention to \emph{sentences} in this logic, i.e., we will assume that
all variables (first-order/function/relation) are quantified.
We will denote by $\vec{x}_i$ (resp. $\vec{y}_i$ ), 
the set of existentially (resp. universally) quantified first order variables
of sort $\sigma_i$, for every $0\leq i\leq k$.

\vspace{-0.4cm}

\paragraph*{\textbf{The problem.}}
The problem we consider is that of deciding satisfiability of \logic sentences
with \emph{background theories} for the background sorts. First we
introduce some concepts.

An \emph{uninterpreted combination of theories (UCT)} over a pure
signature, with $\{\sigma_0, \sigma_1, \ldots, \sigma_k\}$ as the set of sorts, 
is the union of theories $\{T_{\sigma_i}\}_{1 \leq i \leq k}$, where each
$T_{\sigma_i}$ is a theory over signature $\sigma_i$. A sentence
$\phi$ is $\bigcup_{i=1}^k T_{\sigma_i}$-satisfiable if there is a
multi-sorted structure $\Mm$ that satisfies $\phi$ and all the
sentences in $\bigcup_{i=1}^k T_{\sigma_i}$.

The satisfiability problem for \logic with background theories is
the following. Given a UCT $\{T_{\sigma_i}\}_{1\leq i\leq k}$ and a
sentence $\phi \in \logic$, determine if $\phi$ is $\bigcup_{i=1}^k
T_{\sigma_i}$-satisfiable.
We show that this is a decidable problem, and furthermore, there is a
decision procedure that uses a finite number of black-box calls to 
satisfiability solvers of the underlying theories to check
satisfiability of \logic sentences.

For the rest of this paper, for technical convenience, we will assume
that the boolean theory $T_\bool$ is one of the background theories.
This means $\bool \in \Ss$ and the constants
$\top\!:\!\bool, \bot\!:\!\bool \in \Ff$.  The set of sentences in
$T_\bool$ is $\Aa_\bool = \{\top \neq \bot, \forall
(y\!:\!\bool) \cdot (y = \top \lor y = \bot) \}$.  Note that checking
satisfiability of a $\exists^*\forall^*$ sentence over $T_\bool$ is
decidable.

\section{The Decision Procedure for \logic}
\label{sec:dec}
%!TEX root = main.tex

% \vspace{-0.2cm}

In this section we present our decidability result
for sentences over \logic in presence of background theories.
%
%An FO theory $T = (\Sigma_T, \Aa_T)$ is said to be $\exists\forall$-decidable if the problem
%of checking whether sentences of the form 
%$\psi = \exists \vec{x}\, \forall \vec{y}\,\phi$ 
%over $\Sigma_T$-satisfiable is $T$-satisfiable (i.e., checking if there is a model that satisfies
%the theory and the sentence) is decidable.
%
Let us first state the main result of this paper.

\begin{theorem}
\label{thm:mainTheorem}
Let $\Sigma$ be a {\strict} signature with foreground sort $\sigma_0$
and background sorts $\sigma_1, \ldots, \sigma_k$.  Let
$\{T_{\sigma_i}\}_{1 \leq i \leq k}$ be a UCT such that, for each $i$,
checking $T_{\sigma_i}$-satisfiability of $\exists^*\forall^*$
first-order sentences is decidable. Then the problem of checking
$\bigcup\limits_{i=1}^k T_{\sigma_i}$-satisfiability of \logic sentences is
decidable.\qed
\end{theorem}

% \vspace{-0.1cm}

We will prove the above theorem by showing that any
given \logic sentence $\phi$ over a \uct~signature $\Sigma$ can be
transformed, using a sequence of satisfiability preserving
transformation steps, to the satisfiability of
$\exists\forall$ first-order formulae over the individual theories.
% $T_{\sigma_1}, \ldots, T_{\sigma_k}$.
%This sequence of transformations remove the second-order quantifications (both existential and universal)
%till we get a formula that's a first order $\exists^*\forall^*$ formula over the \emph{combined} theories,
%and then we reduce the satisfiability of this formula to satisfiability questions over the individual theories.

%\subsection{Decision Procedure}

%
%	\begin{tikzpicture}
%	\matrix (m) [matrix of math nodes,row sep=0em,column sep=1em,minimum width=2em]
%	{
%		~ & \textit{Step 2a} & ~ & ~ \\
%		\textit{Step 1} & ~ & \textit{Step 3} & \textit{Step 4} \\
%		~ & \textit{Step 2b} & ~ & ~ \\};
%	\path[-stealth]
%	(m-2-1) edge  (m-1-2)
%	(m-2-1) edge  (m-3-2)
%	(m-1-2) edge  (m-2-3)
%	(m-3-2) edge  (m-2-3)
%	(m-2-3) edge  (m-2-4);
%
%	\end{tikzpicture}
	
%Step~1 re
%Step~2 can be achieved in \emph{two} ways, both of which use the \emph{small-model property}
%of the $\exists^*\forall^*$ fragment over the \emph{foreground} theory (similar to the key property
%of the EPR logic that gives it a decidable satisfiability problem).
%Step 2(a) eliminates existential functional variables.
%
%we can either eliminate the foreground sort
%	
%transforms the formula to an equisatisfiable 
%the result of which is a first order 
%(i.e., no quantification over relational or functional variables) 
%\logic sentence over some background theories. 
%We then show how to employ $\exists\forall$ solvers 
%for each of the background theories to decide satisfiability of the FO formula
%obtained after Step 3. This is described in Step-4.

We give a brief overview of the sequence of transformations (Steps $1$
through $4$).  In Step 1, we replace the occurrence of every relation
variable $\Rr$ (quantified universally or existentially) of sort $w$
by a function variable $F$ of sort $(w, \bool)$.  Note that doing this
for the outer existentially quantified relation variables keeps us
within the syntactic fragment.
In Step 2, we eliminate function variables that are existentially
quantified. This crucially relies on the \emph{small model property}
for the foreground universe, similar to EPR~\cite{Bernays1928}. 
This process, however, adds both existential first-order variables 
and universally quantified function variables.
In Step 3, we eliminate the universally quantified function variables
using a standard Ackermann reduction~\cite{pnueli2002small}, which adds more universally
quantified first-order variables.

The above steps result in a first-order $\exists^*\forall^*$ sentence
over the combined background theories, and the empty theory for the
foreground sort.  In Step 4, we show that the satisfiability of such a
formula can be reduced to a finite number of satisfiability queries of
$\exists^*\forall^*$ sentences over \emph{individual} theories.  
% Since
% the latter are decidable, we get our decision procedure. This last
% step is similar to Nelson-Oppen combination in the sense that the
% individual theories have to agree on a contract~\footnote{The notion
% of ``contract'' used here is different from ``arrangements'' used in
% Nelson-Oppen.}. Since the number of such contracts can be exponential,
% the last step causes a blow-up.
% % ; the previous steps only cause a
% % polynomial blow-up. 
% Na\"{i}vely enumerating these agreements is not a
% good idea in practice, and we discuss this issue later in
% Section~\ref{sec:conclusion}.

% 
%recall that these function variables map
%tuples in the foreground sort to elements in some background sort.
%This step crucially relies on a small-model property (stated in Lemma~\ref{lem:small_model}).
%In step-3, we obtain $\step{3}{\phi}$ by eliminating 
%the occurrence of all function variables
%from $\step{2}{\phi}$; such variables are the only
%second order variables in $\step{2}{\phi}$.

% \vspace{-0.4cm}

\subsection*{Step 1 : Eliminating relation variables}

% \vspace{-0.1cm}

The idea here is to introduce, for every relational variable $\Rr$ (with type $w$), a
function variable $f^\Rr$ (with type $(w, \sigma_\bool)$) that corresponds to the characteristic
function of $\Rr$. 
% and replace relational atoms (using $X$) by equality atoms,
%where the equality is between terms, on of which is a boolean constant $\top$,
%and the other is an application of the newly introduced function $f_X$
%to the same arguments as that of the relational atom.

% We first augment $\Sigma = (\Ss, \Ff, \Rf, \varfamily, \funvarfamily, \relvarfamily)$
% with a new sort $\bool$.

% Let $\hat{\Ss} = \Ss \cup \{\bool\}$, $\hat{\Ff} = \Ff \cup \{F_\bool =\{\top, \bot\}\}$ and
% $V_\bool$ be an infinite set of variables of sort $\bool$. 
% Let $\hat{\Sigma} = (\hat{\Ss}, \hat{\Ff}, \Rf, \varfamily \cup V_\bool, \funvarfamily, \relvarfamily)$.

% Let $\Sigma_\bool = \proj{\hat{\Sigma}}{\bool}$, 
% $\Aa_\bool = \{\top \neq \bot, \forall (y\!:\!\bool) \cdot (y = \top \lor y = \bot) \}$.
% Then, the theory of booleans is $T_\bool = (\Sigma_\bool, \Aa_\bool)$.
% \ucomment{Maybe push this `bool-completion' to the previous section}

% We now present the transformation to eliminate relational variables.
Let $\phi$ be \logic formula over $\Sigma$.  We will transform $\phi$
to an \logic formula $\step{1}{\phi}$ over the same signature $\Sigma$.
Every occurrence of an atom of the form
$\Rr(t_1\!:\!\sigma_{i_1}, \ldots, t_k\!:\!\sigma_{i_k})$ in $\phi$,
is replaced by $f^{\Rr}(t_1\!:\!\sigma_{i_1}, \ldots,
t_k\!:\!\sigma_{i_k}) = \top$ in $\step{1}{\phi}$.  Further, every
quantification $Q (\Rr:w)$ is replaced by $Q (f^{\Rr}:w,\bool)$, where
$Q \in\{\forall, \exists\}$.
Thus, the resultant formula $\step{1}{\phi}$ has no relation variables.
Further, it is a \logic formula, since the types of the newly introduced 
existentially quantified function variables are of the form $(\sigma_0^+, \sigma_\bool)$.
% Let the resulting formula be $\step{1}{\phi}$.
%
The correctness of the above transformation is captured by the following lemma.

% \vspace{-0.3cm}

\begin{lemma}
$\phi$ is $\bigcup\limits_{i=1}^k T_{\sigma_i}$-satisfiable iff 
$\step{1}{\phi}$ is $\bigcup\limits_{i=1}^k T_{\sigma_i}$-satisfiable.
\end{lemma}

% \vspace{-0.5cm}

\subsection*{Step 2: Eliminating existentially quantified function variables}

We first note a small-model property with respect to the foreground
sort for $\logic$ sentences.  This property crucially relies on the
fact that existentially quantified function variables do not have
their ranges over the foreground sort.

% \vspace{-0.1cm}

\begin{lemma}[Small-model property for $\sigma_0$]
\label{lem:small_model}
Let $\phi$ be an \logic sentence with foreground sort $\sigma_0$ and
background sorts $\sigma_1, \ldots \sigma_k$.  Let $n$ be the number
of existentially quantified first-order variables of sort $\sigma_0$ in $\phi$.
Then, $\phi$ is $\cup_{i=1}^k T_{\sigma_i}$-satisfiable iff there is a
structure $\Mm = (\{M_{\sigma_i}\}_{i=0}^k, \Ii)$, such that
$|M_{\sigma_0}| \leq n$, $\Mm \models \cup_{i=1}^k T_{\sigma_i}$ and $\Mm \models \phi$.
\end{lemma}

\begin{proof}[Proof (Sketch).]
We present the more interesting direction here.
Consider a model $\Mm = (\Uu, \Ii)$ such that 
$\Mm \models \bigcup\limits_{i=1}^k T_{\sigma_i}$ and $\Mm \models \phi$.
Let $\Ii_{\exists}$ be the interpretation function
that \emph{extends} $\Ii$ so that $(\Uu, \Ii_\exists) \models \varphi$,
where $\varphi$ is the inner universally quantified subformula of $\phi$.
Let $U = \{\Ii_\exists(x) \in M_{\sigma_0}\,|\, x \in \vec{x}_0\}$
be the restriction of the foreground universe to the
interpretations of the variables $\vec{x}_0$.
Clearly, $|U| \leq |\vec{x}_0|$.

Let us first show that $(\proj{\Uu}{U}, \proj{\Ii_\exists}{U}) \models \varphi$.
For this, first see that for every extension 
$\Ii_{\exists\forall}$ of $\Ii_\exists$ with interpretations of all the universal FO variables, 
we must have have $(\Uu, \Ii_{\exists\forall}) \models \psi$, where $\psi$ is the quantifier free part of $\varphi$ (and thus also of $\phi$).
% \begin{align}
% \label{eqn:eq1}
% (\Uu, \Ii_{\exists\forall}) \models \psi.
% \end{align}
% Here $\psi$ is the quantifier free part of $\varphi$ (and thus also of $\phi$).
Now, clearly $(\Uu, \Ii_{\exists\forall}) \models \psi$ must also hold
for those extensions $\Ii_{\exists\forall}^U$
which map all universal variables in $\vec{y}_0$
to the set $U$ and
maps all universally quantified function variables of range sort $\sigma_0$
to function interpretations whose ranges are limited to the set $U$.

Thus, it must also be the case that when we restrict the
universe $M_{\sigma_0}$ to the set $U$,
we have that $(\proj{\Uu}{U}, \proj{\Ii_\exists}{U}) \models \forall*\psi$.
This is because every universal extension $\Ii'$ 
of $\proj{\Ii_\exists}{U}$ is also a projection of one of
these $\Ii_{\exists\forall}^U$ interpretations.
\end{proof}

% \vspace{-0.3cm}

The proof of the above statement 
shows that if there is a model that satisfies $\phi$ (in Lemma~\ref{lem:small_model}), 
then there is a model that satisfies $\phi$ and in which the foreground 
universe contains \emph{only} elements that are
interpretations of the first-order variables $\vec{x}_0$ over the
foreground sort (and hence bounded).  Consequently, instead of
existentially quantifying over a function $F$ (of arity $r$) from the
foreground sort $\sigma_0$ to some background sort $\sigma_i$, we can
instead quantify over first-order variables $\vec{x}^F$ of sort
$\sigma_i$ that capture the image of these functions for each $r$-ary
combination of $\vec{x}_0$.
% 
%
%Now, the only existentially quantified functional variables
%that are allowed in \logic formulae map the foreground sort to one of the
%background sorts.
%Since, the foreground sort $\sigma_0$ admits a small-model property,
%we will replace each existential function variable of arity $m$, 
%by $n^m$ FO variables representing the value of the function variable on
%each combination of the existentially quantified first-order variables
%of sort $\sigma_0$.

% An existentially quantified functional variable $F$ can have
% universally quantified $\vec{y}_0$-variables as arguments. This leads
% to a technical problem.  We handle this challenge by introducing a new
% universally quantified \emph{function} variable $G$ of the same type
% as $F$, and demand that whenever the $G$ is equivalent to the function
% modeled by the newly introduced variables $\vec{x}^F$, the formula,
% with $F$ replaced by $G$, holds.  In order to preserve satisfiability,
% we must also check the formula only when the $\vec{y}_0$ variables map
% to an interpretation of the $\vec{x}_0$ variables.

%We first give the transformation, and postpone the technical subtlety
%involved to the proof of correctness of the transformation.

Let $\step{1}{\phi}$ be the \logic sentence over $\Sigma$ obtained after eliminating relation variables.
Let $\step{1}{\psi}$ be the quantifier free part (also known as the \emph{matrix}) of $\step{1}{\phi}$.
%  and let $\vec{x}_0$ and $\vec{y}_0$
% be respectively the existentially and universally 
% quantified variables of the foreground sort $\sigma_0$.
%
Let $\mathcal{T}_0$ be the set of all sub-terms of sort $\sigma_0$ that occur in $\step{1}{\psi}$.
Note that, $\mathcal{T}_0$ also includes the set of variables and $\vec{x}_0$ and $\vec{y}_0$.
Now, define\footnote[8]{This is a correction from the previous version of the paper~\cite{madhusudan_et_al:LIPIcs:2018:9698}.} %$\widetilde{\psi} = \psi_{\textsf{restrict}} \land \psi$, where the formula $\psi_{\textsf{restrict}}$ is
\vspace{-0.2cm}
\[
\widetilde{\psi}\equiv \psi_{\textsf{restrict}} \land \step{1}{\psi}, \text{ where, }\psi_{\textsf{restrict}}
\equiv \bigwedge\limits_{t \in \mathcal{T}_0} \big(\bigvee\limits_{x \in \vec{x}_0} t = x\big).
\]
\vspace{-0.1cm}
Let $\widetilde{\phi}$ the sentence obtained by replacing the matrix $\step{1}{\psi}$ in $\step{1}{\phi}$,
by $\widetilde{\psi}$.
Then, the correctness of this transformation is noted below.

% \vspace{-0.3cm}
% 
\begin{lemma}
$\step{1}{\phi}$ is $\bigcup\limits_{i=1}^k T_{\sigma_i}$-satisfiable iff 
$\widetilde{\phi}$ is $\bigcup\limits_{i=1}^k T_{\sigma_i}$-satisfiable.
\end{lemma}

% \vspace{-0.3cm}

%The above tranformation essentially ensures that the universal variables
%$\vec{y}_0$ only take values from the sub-model induced by the existential variables $\vec{x}_0$.

We now eliminate the existentially quantified function variables 
in $\widetilde{\phi}$, one by one.
Let $\widetilde{\phi} = (\exists F\!:\!\sigma_0^m, \sigma) \exists^*\forall^* \, \widetilde{\psi}$, 
where $\sigma$ is a background sort.
For every $m$-tuple $t = (t[1], \ldots, t[m])$ over the set $\vec{x}_0$, we introduce 
a variable $x_t^F$ of sort $\sigma$. Let $\vec{x}^F$ be the set of all such $n^m$ variables,
where $n = |\vec{x}_0|$ is the number of existential first order variables of sort $\sigma_0$
 in $\widetilde{\phi}$.
% $\sigma$, say $x^F_{x_{i_1}, \ldots x_{i_m}}$  for every $m$-tuple of integers from $1$ through $n$.
% We will succinctly represent the ordered set of these variables as $\vec{x}^X$.
Next, we introduce a fresh function variable $G^F$ of sort $\sigma_0^m, \sigma$,
and quantify it universally.
$G^F$ will be used to \emph{emulate} the function $F$.
Let us define 
%
% \vspace{-0.5cm}
%
\[
\step{2}{\psi} \equiv (\forall G^F:\sigma_0^m, \sigma) \big( \psi_{\textsf{emulate}} {\implies} \bar{\psi}\big)
\]
where, $\psi_{\textsf{emulate}} \equiv \bigwedge\limits_{t \in \vec{x}_0^m} \big( G^F(t[1], \ldots, t[m]) = x^F_t \big)$
% \[
% \psi_{\textsf{emulate}} \equiv \bigwedge\limits_{t \in \vec{x}_0^m} \big( G^F(t[1], \ldots, t[m]) = x^F_t \big)
% \]
and $\bar{\psi}$ is obtained by replacing all occurrences of $F$ in $\widetilde{\psi}$
by $G^F$.
Now define $\step{2}{\phi}$ to be the sentence
%
% \vspace{-0.2cm}
%
\[
\step{2}{\phi} \equiv (\exists \vec{x}^F:\sigma) \exists^*\forall^* (\forall G^F:\sigma_0^m, \sigma) \, \step{2}{\psi}.
\]
The following lemma states the correctness guarantee of this transformation.
%
% \vspace{-0.3cm}
\begin{lemma}
$\step{2}{\phi}$ is $\bigcup\limits_{i=1}^k T_{\sigma_i}$-satisfiable iff 
$\step{1}{\phi}$ is $\bigcup\limits_{i=1}^k T_{\sigma_i}$-satisfiable.
\end{lemma}

%\subsection*{Step 2b : Eliminating the foreground sort}
%
%\ucomment{TODO}
%The small-model property stated in Lemma~\ref{lem:small_model}
%can be exploited to completely eliminate symbols
%of sort $\sigma_0$ from the formula $\hat{\phi}$.

% \vspace{-0.5cm}

\subsection*{Step 3 : Eliminating universal function variables}

% The recipe here is to perform Ackermann reduction~\cite{ackermann54}.
% For every universally quantified function variable $F\!:\! w, \sigma$ of arity $m$,
% and for every term occurrence of a term $t$ of the form $F(t_1, t_2, \ldots t_m)$,
% we will replace $t$ by a fresh first order variable of the range sort $\sigma$.
% We will then augment the quantifier free part with
% a formula that ensures consistency of this transformation.
The recipe here is to perform Ackermann reduction~\cite{ackermann54} for every universally quantified function variable.
% Let $F$ be an $m$-ary universally quantified function variable in $\step{2}{\phi}$.
% For every occurrence of a term of the form $F(t_1, t_2, \ldots t_m)$ in $\step{2}{\phi}$, 
% we introduce a fresh variable of its range sort.  
% We then augment the quantifier free part with a formula that ensures
% consistency of this transformation.

Let $\step{2}{\phi} \equiv \exists^*\forall^* (\forall
F:w, \sigma) \, \step{2}{\psi}$, where $\step{2}{\psi}$ is the
quantifier free part of $\step{2}{\phi}$, and let $|w| = m$.
For every term $t$ of the form $F(t_1, \ldots, t_m)$ in $\step{2}{\psi}$,
we introduce a fresh first order variable $y^F_{(t_1, t_2, \ldots, t_m)}$ 
of sort $\sigma$, and replace every occurrence of the term $t$ in $\step{2}{\psi}$
with $y^F_{(t_1, t_2, \ldots, t_m)}$.
Let $\widehat{\psi}$ be the resulting quantifier free formula.
Let $\vec{y}^F$ be the collection of all the 
newly introduced variables.
% 
% \noindent 
Let us now define $\step{3}{\psi} \equiv \big(\psi_{\textsf{ack}} \implies \widehat{\psi}$\big). Here, $\psi_{\textsf{ack}} \equiv 
\bigwedge\limits_{y^F_t, y^F_{t'} \in \vec{y}^F} \quad
\Big[(\bigwedge\limits_{j=1}^m t_j = t'_j) \implies (y^F_t =  y^F_{t'})\Big] $
%
% \vspace{-0.3cm}
%
% \[
% \psi_{\textsf{ack}} \equiv 
% \bigwedge\limits_{y^F_t, y^F_{t'} \in \vec{y}^F} \quad
% \Big[(\bigwedge\limits_{j=1}^m t_j = t'_j) \implies (y^F_t =  y^F_{t'})\Big] 
% \]
where, $t = F(t_1, \ldots t_m), t' = F(t'_1, \ldots, t'_m)$.
Then, the transformed formula $\step{3}{\phi} \equiv \exists^*\forall^*
(\forall \vec{y}^F\!:\!\sigma) \, \step{3}{\psi}$ is correct:
\begin{lemma}
$\step{2}{\phi}$ is $\bigcup\limits_{i=1}^k T_{\sigma_i}$-satisfiable iff 
$\step{3}{\phi}$ is $\bigcup\limits_{i=1}^k T_{\sigma_i}$-satisfiable.
\end{lemma}

% \vspace{-0.35cm}

\subsection*{Step-4: Decomposition and black box calls to $\exists^*\forall^*$ Theory solvers}

The \logic sentence $\step{3}{\phi}$ obtained after the sequence of
steps 1 through 3 is a first order $\exists^*\forall^*$ sentence over
$\Sigma$.  This sentence, however, may possibly contain occurrences of
variables of the foreground sort $\sigma_0$. Intuitively, the objective of
this step is to decompose $\step{3}{\phi}$ into $\exists^*\forall^*$
sentences, one for each sort, and then use decision procedures
for the respective theories to decide satisfiability of the decomposed
(single sorted) sentences.  Since such a decomposition can result into
$\exists^*\forall^*$ sentences over the foreground sort, we must
ensure that there is indeed a decision procedure to achieve this.  For
this purpose, let us define $T_{\sigma_0}$ be the empty theory
% \footnote{Recall that the signature is assumed to be \emph{pure} and thus there are no relation and function symbols of sort $\sigma_0$} 
(that
is $\Aa_\sigma = \varnothing$).  Checking satisfiability of
$\exists^*\forall^*$ sentences over $T_{\sigma_0}$ is decidable.
Also, satisfiability is preserved in the presence of $T_{\sigma_0}$ in
the following sense.

% \vspace{-0.2cm}

\begin{lemma}
$\step{3}{\phi}$ is $\bigcup\limits_{i=1}^k T_{\sigma_i}$-satisfiable iff 
$\step{3}{\phi}$ is $\bigcup\limits_{i=0}^k T_{\sigma_i}$-satisfiable.
\end{lemma}

We first transform the quantifier free part $\step{3}{\psi}$  of $\step{3}{\phi}$
into an equivalent CNF formula $\psi_{\textsf{CNF}}$. Let $\phi_{\textsf{CNF}}$
be obtained by replacing $\step{3}{\psi}$ by $\psi_{\textsf{CNF}}$.
% Let $\psi_{\textsf{CNF}}$ be a conjunction of $r$ clauses $\psi_j$, $1\leq j\leq r$.
Let $\phi_{\textsf{CNF}} \equiv  \exists^*\forall^* \psi_{\textsf{CNF}}$, where $\psi_{\textsf{CNF}} \equiv \bigwedge\limits_{i=1}^r \psi_i$ and each $\psi_i$ is a disjunction of atoms.
%
% \vspace{-0.3cm}
% \noindent
% where $\psi_i$ is a disjunction of atoms.
% 
Since $\phi_{\textsf{CNF}}$ is a first order
formula over a \strict~signature,
all atoms are either of the form $R(\cdots)$
or $t = t'$ (with possibly a leading negation).
Now, equality atoms are restricted to
terms of the same sort.
Also since $\Sigma$ is \strict, the argument terms of
all relation applications have the same sort.
This means, for every atom $\alpha$, there is a unique associated
sort $\sigma \in \Ss$, which we will denote
by $\textsf{sort}(\alpha)$.

For a clause $\psi_i$ in $\psi_\textsf{CNF}$, let
$\textsf{atoms}(\psi_i)$ be the set of atoms in $\psi_i$.  Let
$\textsf{atoms}^\sigma(\psi_i)
= \{ \alpha \in \textsf{atoms}(\psi_i) \, |\, \textsf{sort}(\alpha)
= \sigma\}$, and let $\psi_{i}^\sigma
\equiv \bigvee\limits_{\alpha \in \textsf{atoms}^\sigma(\psi_i)} \alpha$.
Then, we have the identity $\psi_{\textsf{CNF}}
\equiv \bigwedge\limits_{j=1}^r \bigvee\limits_{\sigma \in \Ss} \psi_j^\sigma$.
We now state our decomposition lemma.

% \vspace{-0.3cm}

\begin{lemma}
\label{lem:decompose}
$\phi_\textsf{CNF}$ is $\bigcup\limits_{i=0}^k T_{\sigma_i}$-satisfiable iff there is a mapping $L : \{1, \ldots, r\} \to \Ss$
such that for each $0\leq i\leq k$ , the formula $\phi^L_i \equiv (\exists \vec{x}_i : \sigma_i)(\forall \vec{y}_i : \sigma_i) \bigwedge\limits_{j \in L^{-1}(\sigma_i)} \psi_j^{\sigma_i}$ is $T_{\sigma_i}$-satisfiable.
\end{lemma}

\begin{proof}[Proof (Sketch).]
We present the more interesting direction here.
Let $\phi_{\textsf{Skolem}}$ be an equi-satisfiable Skolem norm form of $\phi_{\textsf{CNF}}$. That is,  
$\phi_{\textsf{Skolem}} = \forall^* \psi_{\textsf{Skolem}}$, where
$\psi_{\textsf{Skolem}}$ is obtained from $\psi_{\textsf{CNF}}$
by replacing all existential variables $\vec{x}_0, \vec{x}_1 \ldots, \vec{x}_k$
by \emph{Skolem} constants.
% That is, propositional (or boolean) structure of $\psi_{\textsf{CNF}}$ and
% $\psi_{\textsf{Skolem}}$, and 
We will use the same notation $\psi_i$
for the $i^{\text{th}}$ clause of $\psi_{\textsf{Skolem}}$.
Then, consider a structure $\Mm$ such that $\Mm \models \bigcup\limits_{i=0}^k T_{\sigma_i}$
and $\Mm \models \phi_{\textsf{Skolem}}$.
% First note that the individual universes $M_{\sigma_i}$ ($0\leq i\leq k$)
% and the corresponding projections $\proj{\Ii}{M_{\sigma_i}}$ of 
% the interpretation function $\Ii$ satisfy the individual theories :
% $(M_{\sigma_i}, \proj{\Ii}{M_{\sigma_i}}) \models T_{\sigma_i}$.
Now, suppose, on the contrary, that there is a
clause $\psi_j$ such that for every sort $\sigma_i$,
we have $\Mm \not\models \forall (\vec{y}_i\!:\!\sigma_i)\psi_j$.
This means, for every sort $\sigma_i$, there is a 
interpretation $\Ii_i$ (that extends $\Ii$ with valuations of $\vec{y}_i$),
such that either $\Ii_i$ leads to falsity of $T_{\sigma_i}$
or the clause $\psi_j$.
Let $c^{\sigma_i}_1, c^{\sigma_i}_2, \ldots c^{\sigma_i}_{|\vec{y}_i|}$
be the values assigned to the universal variables $\vec{y}_i$
in $\Ii_i$.
Then, construct an interpretation $\Ii'$ by
extending $\Ii$ with the variables $\vec{y}_i$
interpreted with $c^{\sigma_i}$'s .
This interpretation $\Ii'$ can be shown to either violate
one of the theory axioms or the formula $\psi_j$.
In either case, we have a contradiction.
\end{proof}

The \emph{contract} $L$ above identifies, for each clause $\psi_j$,
one sort $\sigma_i$ such that the restriction $\psi_j^{\sigma_i}$ of $\psi_j$ to $\sigma_i$
can be set to true. Thus, in order to decide
satisfiability of $\phi_\textsf{CNF}$, a straightforward decision
procedure involves enumerating all contracts, $L \in [\{1, \ldots,
r\} \to \Ss]$.  For each contract $L$ and for each sort $\sigma_i$, we
construct the sentence $\phi^L_i$, and make a black-box call to the
$\exists^*\forall^*$ theory solver for $T_{\sigma_i}$.  If there is a
contract $L$ for which each of these calls return `SATISFIABLE', then
$\phi_\textsf{CNF}$ (and thus, the original formula $\phi$) is
satisfiable.  Otherwise, $\phi$ is unsatisfiable.
% \ucomment{However, as stated in Section~\ref{}, such a 
% naive enumeration is exponential.}

\section{Undecidability Results}

\label{sec:undec}
%!TEX root = main.tex
 
The logic that we have defined was carefully chosen to avoid
undecidability of the satisfiability problem.  We now show that
natural generalizations or removal of restrictions in our logic
renders the satisfiability problem undecidable.  We believe our
results are hence not simple to generalize any further.
 
One restriction that we have is that the functions that are existentially quantified
cannot have $\sigma_0$ as their range sort.
A related restriction is that the \emph{universal} quantification block
quantifies all uninterpreted function symbols, as otherwise they must be
existentially quantified on the outside block.
 
Let us now consider the fragment of logic where formulas are of the form 
$(\exists \vec{x}_0) \, (\exists \vec{F}) (\forall \vec{y}_0) \, \psi$ where in fact we do not
even have any background theory. Since the formula is over a single
sort, we have dropped the sort annotations on the variables. It is not
hard to see that this logic is undecidable.
 % 
 % \vspace{-0.2cm}
 % 
\begin{theorem}
\label{thm:undec1}
Consider signature with a single sort $\sigma_0$ (and no background
sorts). 
The satisfiability problem for sentences of the following form is undecidable. 
%
% \vspace{-0.2cm}
%
\[
(\exists \vec{x}_0) \, (\exists \vec{F}) (\forall \vec{y}_0) \, \psi
\]
\end{theorem}

\begin{proof}[Proof (Sketch)] 
We can show this as a mild modification of standard proofs of the
undecidability of first-order logic. We can existentially quantify
over a variable $Zero$ and a function $\textit{succ}$, demand that for
any element $y$, $\textit{succ}(y)$ is not $Zero$, and for every $y,
y'$, if $\textit{succ}(y) = \textit{succ}(y')$, then $y=y'$. This
establishes an infinite model with distinct elements
$\textit{succ}^n(Zero)$, for every $n\geq 0$.  We can then proceed to
encode the problem of \emph{non-halting} of a 2-counter machine using
a relation $R(t,q,c_1,c_2)$, which stands for the 2CM is in state $q$
at time $t$ with counters $c_1$ and $c_2$, respectively. It is easy to
see that all this can be done using only universal quantification (the
relation $R$ can be modeled as a function easily).
\end{proof}
 %
 % \vspace{-0.2cm}
 %
The theorem above has a simple proof, but the theorem is not new; in fact, even
more restrictive logics are known to be undecidable
(see~\cite{borger2001classical}).
 
% Another important restriction that we have is that the foreground sort
% and the various background sorts are \emph{pariwise disjoint}. 
% This requirement is also not negotiable if decidability is desired, as it
% is easy to show the following result.
% Once again we have dropped sort annotations, since we only have a single sort.
Based on Theorem~\ref{thm:undec1}, when the only sort we have is the foreground sort,
quantifying functions existentially results in undecidability of the satisfiability
problem. On the other hand, if we allow only relation symbols (over the foreground sort) 
to be existentially quantified, the satisfiability problem becomes decidable~\cite{Bernays1928}.
However, for certain theories different from the empty theory (such as Presburger arithmetic), 
existential quantification over relation symbols (without any quantified function symbols) 
can lead to undecidability, 
despite the decidability of checking satisfiability of the entire first order fragment. 
This is the content of Theorem~\ref{thm:undec2}
and sheds light on the importance of the restriction that the foreground sort
is disjoint from the various background sorts in~\logic.
%
%
% \vspace{-0.1cm}
%
\begin{theorem}
\label{thm:undec2}
Consider a signature with a single sort $\sigma_1$ and let
$T_{\sigma_1}$ be the theory of Presburger arithmetic. The
satisfiability problem is decidable for sentences of the form
% \vspace{-0.2cm}
\[
(\exists \vec{x}_1) \, (\exists \vec{R}) \, (\forall \vec{y}_1) \, \psi
\]
%
% \vspace{-0.4cm}
% \noindent
% is undecidable.
\end{theorem}

\begin{proof}[Proof (Sketch)]
We can use a similar proof as the theorem above, except now that we
use the successor function available in Presburger arithmetic. We can
again reduce non-halting of Turing machines (or 2-counter machines) to
satisfiability of such formulas.
\end{proof}
 % 
 % \vspace{-0.1cm}
 % 
Stepping further back, there are very few subclasses of first-order
logic with equality that have a decidable satisfiability problem, and
the only standard class that admits $\exists^* \forall^*$ prefixes is
the Bernays-Sch\"onfinkel-Ramsey class
(see~\cite{Bernays1928}). 
Our results can be seen as an
extension of this class with background theories, where the background
theories admit locally a decidable satisfiability problem for the
$\exists^*\forall^*$ fragment.
% \vspace{-0.2cm}

\section{Applications to Synthesis}

\label{sec:exp}
%!TEX root = main.tex

\subsection{Synthesis: Validity or Satisfiability?}

Though we argued in Section~\ref{sec:example} that synthesis problems
can be modeled using satisfiability of \logic sentences, there is
one subtlety that we would like to highlight. In synthesis problems,
we are asked to find an expression such that the expression satisfies
a specification expressed as a formula in some logic. Assuming the specification is
modeled as a universally quantified formula over background theories,
we would like to know if $\forall y \varphi(e, y)$ holds for the
synthesized expression $e$.  However, in a logical setting, we have to
qualify what ``holds'' means; the most natural way of phrasing this is
that $\forall y \varphi(e, y)$ is valid over the underlying background
theories, i.e., holds in all models that satisfy the background
theories. However, the existential block that models the existence of
an expression is clearly best seen as a satisfiability problem, as it
asks whether there is \emph{some foreground model} that captures an
expression. Requiring that it holds in \emph{all} foreground models
(including those that might have only one element) would be
unreasonable.

To summarize, the synthesis problem is most naturally modeled as a
logical problem where we ask whether there is \emph{some} foreground
model (capturing a program expression) such that \emph{all} 
background models, that satisfy their
respective background theories, also satisfy the quantifier free formula 
expressing that the synthesized expression satisfies the
specification. This is, strictly speaking, neither a satisfiability
problem nor a validity problem!

We resolve this by considering only \emph{complete and consistent}
background theories. Hence validity of a formula under a background
theory $T$ is \emph{equivalent} to $T$-satisfiability. Consequently,
synthesis problems using such theories can be seen as asking whether
there is a foreground universe (modeling the expression to be
synthesized) \emph{and} some background models where the specification
holds for the expression.  We can hence model synthesis purely as a
satisfiability problem of \logicWithComma as described in Section~2.
 
Many of the background theories used in verification/synthesis and SMT
solvers are complete theories (like Presburger arithmetic, FOL over
reals, etc.).  One incomplete theory often used in verification is the
theory of \emph{uninterpreted functions}.  However, in this case,
notice that since the functions over this sort are uninterpreted,
validity of formulas can be modeled using a \emph{universal
  quantification} over functions, which is supported in \logic\!! The only
other adjustment is to ensure that this background theory has only
infinite models (we can choose this background theory to be the theory
of $(\mathbb{N}, =)$, which has a decidable satisfiability
problem). Various scenarios such as modeling pointers in heaps,
arrays, etc., can be naturally formulated using uninterpreted functions
over this domain.

The second issue in modeling synthesis problems as satisfiability
problems for \logic is that in synthesis, we need to construct the
expression, rather than just know one exists. It is easy to see that
if the individual background theory solvers support finding concrete
values for the existentially quantified variables, then we can pull
back these values across our reductions to give the values of the
existentially quantified first-order variables (over all sorts), the
existentially quantified function variables as well as the
existentially quantified relation variables, from which the expression
to be synthesized can be constructed.

\vspace{-0.4cm}

\subsection{Evaluation}

% \vspace{-0.1cm}

We  illustrate the applicability of our result for
solving synthesis problems.

\medskip
\noindent{\bf Synthesis of recursive programs involving lists.}

We model the problem of synthesizing
\emph{recursive} programs with lists, that will
meet a pre/post contract $C$ assuming that recursive calls on smaller
data-structures satisfy the same contract $C$. Though the programs we
seek are recursive, we can model certain classes of programs using straight-line programs.

To see this, let us take the example of synthesizing a program that finds a
particular key in a linked list (\textsf{list-find}). We can instead ask whether there is a
straight-line program which takes an additional input which models the
return value of a possible recursive call made on the tail of the
list. The straight-line program must then work on the head of the list
and this additional input (which is assumed to satisfy the contract
$C$) to produce an output that meets the same contract $C$.

For this problem, we modeled the program to be synthesized using
existential quantification (over a grammar that generates bounded
length programs) as described in Section~\ref{sec:example}. The
pointer \texttt{next} and recursive data structures \texttt{list,
  lseg} in the verification condition were modeled using
\emph{universal quantification} over function variables and relation
variables, respectively.  Moreover, in order to have a tractable
verification condition, we used the technique of \emph{natural
  proofs}~\cite{natproof17,pek2014natural,qiu2013natural} that soundly formulates the condition
in a decidable theory.
% \ucomment{Though we do not have an implementation of our decision procedure, we
% did several of our steps manually, and finally encoded the problem in
% the {\sc SyGuS} format~\cite{sygus} and used an off-the-shelf
% enumerative counter-example guided synthesis (CEGIS) solver.}
We used \textsf{z3}~\cite{DeMoura:2008:ZES:1792734.1792766} to ackermanize the universally quantified functions/relations (\texttt{lseg}, \texttt{list} and \texttt{next}).
We encoded the resulting formula as a synthesis problem
in the \textsc{SyGuS} format~\cite{sygus} and used an off-the-shelf
enumerative counter-example guided synthesis (CEGIS) solver.
A program was synthesized within $1$s, which was
manually verified to be correct.

We also encoded other problems involving lists :
\textsf{list-length} (calculating the length of a list),
\textsf{list-sum} (computing sum of the keys in a list),
\textsf{list-sorted} (checking if the sequence of keys in the list is sorted)
and \textsf{list-count-occurrence} (counting the number of occurrences of a key in the list),
using a CEGIS solver, and report the running times and the number of programs explored
in Table~\ref{tab:table_exp}.

We are convinced that \logic can handle recursive program synthesis (of bounded size)
against separation logics specifications expressed using natural
proofs (as in~\cite{pek2014natural}).

\medskip
\noindent{\bf Synthesis of straight-line programs equivalent to given recursive programs.}

In the second class of examples, we turn to synthesizing straight-line
programs given a recursive function as their specification. For
example, consider Knuth's generalization of the recursive McCarthy 91
function:
\vspace{-0.2cm}
\[
M(n) = \left\{ \begin{array}{ll}
       n-b & \mbox{if } n> a \\
       M^c(n+d)) & \mbox{otherwise}
               \end{array} \right.
\]
%
% \vspace{-0.15cm}
\noindent
for every integer $n$, and where $(c-1)b < d$.
For the usual McCarthy function, we have $a=100$, $b=10$, $c=2$, and $d=11$.

%!TEX root = main.tex

% Please add the following required packages to your document preamble:
% \usepackage{multirow}
% \usepackage[table,xcdraw]{xcolor}
% If you use beamer only pass "xcolor=table" option, i.e. \documentclass[xcolor=table]{beamer}

\begin{table*}[t!]
\captionsetup{font=small}

\centering
\scalebox{0.9}{

\begin{tabular}{!{\VRule[1pt]}c|c|r!{\VRule[1pt]}}

% \hline
% \specialrule{1pt}{0pt}{0pt}
% 1 & 2 & 3 \\ 
\specialrule{1pt}{0pt}{0pt}

\rowcolor[HTML]{DDDDDD}  ~Program
& \cellcolor[HTML]{DDDDDD} \# Programs Explored
& \cellcolor[HTML]{DDDDDD} Time(s)\\

\rowcolor[HTML]{DDDDDD}
& \cellcolor[HTML]{DDDDDD} in \textsc{SyGuS}
& \cellcolor[HTML]{DDDDDD}\\

\specialrule{1pt}{0pt}{0pt}

~\textsf{list-find}  & $\sim$5k & $0.5$~ \\ \hline
~\textsf{list-length}  & $\sim$40k  & $5$~ \\ \hline
~\textsf{list-sum}  & $\sim$160k & $15$~ \\ \hline
~\textsf{list-sorted}  & $\sim$206k  & $45$~ \\ \hline
~\textsf{list-count-occurrence}  & $\sim$1.3 million  & $134$~ \\
\specialrule{1pt}{0pt}{0pt}

~\textsf{Knuth} : $(a = 100$, $b = 10$, $c = 2$, $d = 11)$ &  - & $2$~ \\ \hline
~\textsf{Knuth} : $(a = ~15$, \,$b = 30$, $c = 3$, $d = 61)$ & - & $6$~ \\ \hline %& TO \\ \hline
~\textsf{Knuth} : $(a = ~~3$, \,$b = 20$, $c = 4$, $d = 62)$ & - & 27~ \\ \hline
~\textsf{Knuth} : $(a = ~~9$, \,$b = 11$, $c = 5$, $d = 45)$ & - & 49~ \\ \hline
~\textsf{Knuth} : $(a = ~99$, \,$b =10$, $c = 6$, $d = 51)$ & - & 224~ \\ \hline

~\textsf{Takeuchi}  & - & 100~ \\

\specialrule{1pt}{0pt}{0pt}

\end{tabular}

}
% \end{adjustbox}
% }
\vspace{0.2cm}
\caption{
Synthesis of list programs and recursive programs. 
}
\vspace{-0.3in}
\label{tab:table_exp}
\end{table*}

Consider the problem of synthesizing an equivalent recursion-free
expression. The programs we consider may have if-then-else statements
of nesting depth 2, with conditionals over linear expressions having
unbounded constants. Existential quantification over the background
arithmetic sort allowed us to model synthesizing these unbounded
constants. Our specification demanded that the value of the expression
for $n$ satisfy the recursive equations given above.

% \ucomment{Though we do not have an implementation of our logic engine}, 
We modeled the foreground sort \emph{inside} arithmetic, and converted
our synthesis problem to a first-order $\exists^*\forall^*$ sentence
over Presburger arithmetic and Booleans. We experimented with several
values for $a,b,c,d$ (with $(c-1)b<d$), and interestingly, solutions
were synthesized only when $(d-(c-1)b)=1$. Given Knuth's result that a
closed form expression involves taking remainder modulo this
expression (and since we did not have the modulo operation in our syntax),
it turns out that simple expressions do not exist otherwise. Also,
whenever the solution was found, it matched the recursion-free
expression given by Knuth (see Theorem~1 in~\cite{knuth1991textbook}). 
% For
% instance, when $a=13$, $b=82$, $c=83$, $d=2$, we found the solution
% $\textit{ite}(x>13, x-82, -68)$ in about 25s. We also verified for the following values of ($a$, $b$, $c$, $d$): $(13, 82, 2, 83)$, $(15, 30, 3, 61)$, $(3, 20, 4, 62)$, $(9, 11,5, 45)$, and $(99, 10, 6, 51)$, which took $2s$, $4s$, $27s$, $49s$, and $224s$, respectively, to solve. All the solutions agreed with Knuth's closed form solution.  
In Table~\ref{tab:table_exp}, we provide the running times
of our implementation on various parameters.
We also compared our implementation with
the popular synthesis tool \textsc{Sketch}~\cite{solar2006combinatorial}
on these examples.
For the purpose of comparison, we used the same template for both
\textsc{Sketch} and our implementation.
Further, since \textsc{Sketch} does not allow encoding integers
with unbounded size (unlike our encoding in integer arithmetic),
we represented these constants, to be synthesized, using bitvectors of size $8$.
\textsc{Sketch} does not return an answer within the set time-limit of 10 minutes
for most of these programs. 

We also modeled the Tak function (by Takeuchi) given by the specification below.
%
% \vspace{-0.5cm}
%
\[
t(x,y,z) = \left\{ \begin{array}{ll}
                     y & \mbox{if } x \leq y\\
                     t(t(x-1,y,z), t(y-1,z,x), t(z-1,x,y)) & \mbox{otherwise}
                   \end{array} \right.
\]
%
% \vspace{-0.1cm}
%
Our implementation synthesized the
program $\mathtt{t(x,y,z)} \mathtt{= ite( x\leq y, y, ite(y \leq z, z, x))}$
in about 100s.
% \vspace{-0.1cm}
% \[
% \mathtt{t(x,y,z)} \mathtt{= ite( x\leq y, y, ite(y \leq z, z, x))}.
% \]
% \vspace{-1cm}

\section{Related Work}

\label{sec:related}
%!TEX root = main.tex

There are several logics known in the literature that can 
express synthesis problems and are decidable. 
The foremost example is the monadic second-order theory over trees, 
which can express Church's synthesis problem~\cite{buchi1969solving} 
and other reactive synthesis problems over \emph{finite data domains}, 
and its decidability (Rabin's theorem~\cite{rabin1969decidability})
is one of the most celebrated theorems in logic that is applicable to computer science. 
Reactive synthesis has been well studied and applied in 
computer science~(see, for example,~\cite{Bloem:2007:IPA:1266366.1266622}).
The work reported in~\cite{madhusudan:LIPIcs:2011:3247} is a tad closer 
to program synthesis as done today, as it synthesizes
syntactically restricted programs with recursion that work on finite domains.

Caulfield et al~\cite{CaulfieldRST15} have considered the decidability 
of \emph{syntax-guided synthesis} (SyGuS) problems, where the synthesized 
expressions are constrained to belong to a grammar (with operators 
that have the usual semantics axiomatized by a standard theory 
such as arithmetic) that satisfy a universally quantified constraint. 
They show that the problem is undecidable in many cases, 
but identify a class that asks for expressions satisfying 
a regular grammar with uninterpreted function theory constraints to be decidable. 
 
The $\exists^*\forall^*$ fragment of pure predicate logic 
(without function symbols) was shown to be decidable by Bernays 
and Sch\"onfinkel (without equality)  and by Ramsey 
(with equality)~\cite{Bernays1928}, and is often called 
Effectively Propositional Reasoning (EPR) class. 
It is one of the few fragments of first-order logic known to be decidable.
The EPR class has been used in program 
verification~\cite{itzhaky2013effectively,padon2016ivy}, and
efficient SMT solvers supporting EPR have been developed~\cite{piskac2008deciding}.
 
The work by~\cite{abadi2010decidable} extends EPR to stratified
typed logics, which has some similarity with our restriction that
the universes of the foreground and background be disjoint.
However, the logic therein does not allow background SMT theories
unlike ours and restricts the communication between 
universally and existentially quantified variables
via equality between existential variables and terms with universally
quantified variables as arguments.
% Further, the decidability results of such extensions
% rely on finite model property of the quantifier free formulae
% of the component pure theories, whereas this is not a restriction for decidability of \logicWithPeriod
In~\cite{Horbach2017}, EPR with simple linear arithmetic (without addition)
is shown to be decidable.

Theory extensions~\cite{localtheory2013} and model theoretic
and syntactic restrictions theoreof~\cite{hierarchical2005} have been explored to
devise decidable fragment for quantified fragments of first order logic.
Here, reasoning in \emph{local} theory extensions of a
base theory can be reduced to the reasoning in the base theory
(possibly with an additional quantification).
Combination of theories which are extensions of a common base theory
can similarly be handled by reducing the reasoning
to a decidable base theory.
Similar ideas have been employed in the context of combinations
of linear arithmetic and the theory of uninterpreted functions
with applications to construct interpolants~\cite{Kapur2006} and 
invariants~\cite{beyer2007} for program verification.
\logic does not require the background theories to be
extensions of a common base theory.

Verification of programs with arrays and heaps
can be modeled using second order quantification over 
the arrays/heaps and quantifier alternation over the elements
of the array/heaps which belong to the theory of Presburger arithmetic. 
While such a logic is, in general, undecidable,
careful syntactic restrictions such as limiting quantifier alternation~\cite{bradley2006}
and \emph{flatness} restrictions~\cite{Alberti2015}.
We do not restrict the syntax of our formulae, but ensure decidability
via careful sort restrictions.
A recent paper~\cite{natproof17} develops sound and complete reasoning for 
a so-called \emph{safe} FO fragment of an uninterpreted combination of theories. However,
the logic is undecidable, in general, and also does not support second-order quantification.
 
The SyGuS format has recently been proposed as a language to express \emph{syntax guided}
synthesis problems, and there have been several synthesis engines developed for various
tracks of SyGuS~\cite{sygus}. However, the syntax typically allows unbounded programs, and hence
the synthesis problem is not decidable. 
% Furthermore, unlike our logic, the grammar does not
% allow arbitrarily large constants (like integers) to be synthesized as part of the program.
In~\cite{decorated17}, the candidate program components
are `decorated' with annotations that represent transformers of the components
in a sound abstract domain.
This reduces the synthesis problem $(\exists^*\forall^*)$ to the search for a proof $(\exists^*\exists^*)$ in 
the abstract domain.

When expressing synthesis problems for programs that manipulate heaps,
we rely on natural-proofs style sound abstraction of the verification conditions.
Natural synthesis~\cite{naturalsynth17} extends this idea to an inductive synthesis procedure.

\vspace{-0.2cm}

\section{Conclusions and Future Work}
\label{sec:conclusion}
%!TEX root = main.tex

The logic \logic defined herein is meant to be a decidable logic for communication between
researchers modeling program synthesis problems and researchers developing efficient logic
solvers. 
Such liaisons have been extremely fruitful in verification, where SMT solvers have
served this purpose. 
We have shown the logic to be decidable and its efficacy in modeling
synthesis problems. However, the decision procedure has several costs that should not be
paid \emph{up front} in any practical synthesis tool. Ways to curb such costs are known
in the literature of building efficient synthesis tools. In particular, searching for
foreground models is similar to EPR where efficient engines have been developed~\cite{piskac2008deciding},
and the search can also be guided by CEGIS-like approaches~\cite{sygus}.
And the exponential blow-up caused by guessing contracts between solvers (in Step~4 of our procedure)
is similar to \emph{arrangements} agreed upon by theories combined using the Nelson-Oppen
method, again for which efficient solvers have been developed. 
Our hope is that researchers
working on logic engines will engineer an efficient decision procedure for \logic that
can solve synthesis problems.

\newpage

\bibliographystyle{splncs03}
\bibliography{references}

\begin{thebibliography}{10}
\providecommand{\url}[1]{\texttt{#1}}
\providecommand{\urlprefix}{URL }

\bibitem{abadi2010decidable}
Abadi, A., Rabinovich, A., Sagiv, M.: Decidable fragments of many-sorted logic.
  In: Proceedings of the 14th International Conference on Logic for
  Programming, Artificial Intelligence and Reasoning. pp. 17--31. LPAR'07,
  Springer-Verlag, Berlin, Heidelberg (2007),
  \url{http://dl.acm.org/citation.cfm?id=1779419.1779423}

\bibitem{ackermann54}
Ackermann, W., Ackermann, F.W.: Solvable cases of the decision problem  (1954)

\bibitem{Alberti2015}
Alberti, F., Ghilardi, S., Sharygina, N.: Decision procedures for flat array
  properties. Journal of Automated Reasoning  54(4),  327--352 (Apr 2015),
  \url{https://doi.org/10.1007/s10817-015-9323-7}

\bibitem{sygus}
Alur, R., Bod{\'{\i}}k, R., Dallal, E., Fisman, D., Garg, P., Juniwal, G.,
  Kress{-}Gazit, H., Madhusudan, P., Martin, M.M.K., Raghothaman, M., Saha, S.,
  Seshia, S.A., Singh, R., Solar{-}Lezama, A., Torlak, E., Udupa, A.:
  Syntax-guided synthesis. In: Dependable Software Systems Engineering, pp.
  1--25. {IOS} Press (2015), \url{https://doi.org/10.3233/978-1-61499-495-4-1}

\bibitem{Bernays1928}
Bernays, P., Sch{\"o}nfinkel, M.: Zum entscheidungsproblem der mathematischen
  logik. Mathematische Annalen  (1928)

\bibitem{beyer2007}
Beyer, D., Henzinger, T.A., Majumdar, R., Rybalchenko, A.: Invariant synthesis
  for combined theories. In: Cook, B., Podelski, A. (eds.) Verification, Model
  Checking, and Abstract Interpretation. pp. 378--394. Springer Berlin
  Heidelberg, Berlin, Heidelberg (2007)

\bibitem{Bloem:2007:IPA:1266366.1266622}
Bloem, R., Galler, S., Jobstmann, B., Piterman, N., Pnueli, A., Weiglhofer, M.:
  Interactive presentation: Automatic hardware synthesis from specifications: A
  case study. In: Proceedings of the Conference on Design, Automation and Test
  in Europe. DATE '07 (2007)

\bibitem{borger2001classical}
B{\"o}rger, E., Gr{\"a}del, E., Gurevich, Y.: The classical decision problem.
  Springer Science \& Business Media (2001)

\bibitem{bradley2006}
Bradley, A.R., Manna, Z., Sipma, H.B.: What's decidable about arrays? In:
  Emerson, E.A., Namjoshi, K.S. (eds.) Verification, Model Checking, and
  Abstract Interpretation. pp. 427--442. Springer Berlin Heidelberg, Berlin,
  Heidelberg (2006)

\bibitem{buchi1969solving}
Buchi, J.R., Landweber, L.H.: Solving sequential conditions by finite-state
  strategies. Transactions of the American Mathematical Society  (1969)

\bibitem{CaulfieldRST15}
Caulfield, B., Rabe, M.N., Seshia, S.A., Tripakis, S.: What's decidable about
  syntax-guided synthesis? CoRR  abs/1510.08393 (2015)

\bibitem{DeMoura:2008:ZES:1792734.1792766}
De~Moura, L., Bj{\o}rner, N.: Z3: An efficient smt solver. In: TACAS (2008)

\bibitem{decorated17}
Gasc{\'o}n, A., Tiwari, A., Carmer, B., Mathur, U.: Look for the proof to find
  the program: Decorated-component-based program synthesis. In: Computer Aided
  Verification (2017)

\bibitem{gulwanidimensions10}
Gulwani, S.: Dimensions in program synthesis. In: Proceedings of the 12th
  International ACM SIGPLAN Symposium on Principles and Practice of Declarative
  Programming. pp. 13--24. PPDP '10, ACM, New York, NY, USA (2010),
  \url{http://doi.acm.org/10.1145/1836089.1836091}

\bibitem{Horbach2017}
Horbach, M., Voigt, M., Weidenbach, C.: On the combination of the
  {Bernays--Sch{\"o}nfinkel--Ramsey} fragment with simple linear integer
  arithmetic. In: Proceedings of the International Conference on Automated
  Deduction. pp. 202--219 (2017)

\bibitem{itzhaky2013effectively}
Itzhaky, S., Banerjee, A., Immerman, N., Nanevski, A., Sagiv, M.:
  Effectively-propositional reasoning about reachability in linked data
  structures. In: International Conference on Computer Aided Verification
  (2013)

\bibitem{jha2010oracle}
Jha, S., Gulwani, S., Seshia, S.A., Tiwari, A.: Oracle-guided component-based
  program synthesis. In: Proceedings of the 32nd ACM/IEEE International
  Conference on Software Engineering-Volume 1. pp. 215--224. ACM (2010)

\bibitem{Kapur2006}
Kapur, D., Majumdar, R., Zarba, C.G.: Interpolation for data structures. In:
  Proceedings of the 14th ACM SIGSOFT International Symposium on Foundations of
  Software Engineering. pp. 105--116. SIGSOFT '06/FSE-14, ACM, New York, NY,
  USA (2006), \url{http://doi.acm.org/10.1145/1181775.1181789}

\bibitem{knuth1991textbook}
Knuth, D.E.: Textbook examples of recursion. Artificial Intelligence and
  Mathematical Theory of Computation: Papers in Honor of John McCarthy  (1991)

\bibitem{natproof17}
L\"{o}ding, C., Madhusudan, P., Pe\~{n}a, L.: Foundations for natural proofs
  and quantifier instantiation. Proc. ACM Program. Lang.  2(POPL),  10:1--10:30
  (Dec 2017), \url{http://doi.acm.org/10.1145/3158098}

\bibitem{madhusudan_et_al:LIPIcs:2018:9698}
Madhusudan, P., Mathur, U., Saha, S., Viswanathan, M.: {A Decidable Fragment of
  Second Order Logic With Applications to Synthesis}. In: Ghica, D., Jung, A.
  (eds.) 27th EACSL Annual Conference on Computer Science Logic (CSL 2018).
  Leibniz International Proceedings in Informatics (LIPIcs), vol. 119, pp.
  31:1--31:19. Schloss Dagstuhl--Leibniz-Zentrum fuer Informatik, Dagstuhl,
  Germany (2018), \url{http://drops.dagstuhl.de/opus/volltexte/2018/9698}

\bibitem{madhusudan:LIPIcs:2011:3247}
Madhusudan, P.: {Synthesizing Reactive Programs}. In: Computer Science Logic
  (CSL'11) - 25th International Workshop/20th Annual Conference of the EACSL
  (2011)

\bibitem{manna1969properties}
Manna, Z., McCarthy, J.: Properties of programs and partial function logic.
  Tech. rep. (1969)

\bibitem{no79}
Nelson, G., Oppen, D.C.: Simplification by cooperating decision procedures. ACM
  Transactions on Programming Languages and Systems (TOPLAS)  1(2),  245--257
  (1979)

\bibitem{padon2016ivy}
Padon, O., McMillan, K.L., Panda, A., Sagiv, M., Shoham, S.: Ivy: safety
  verification by interactive generalization. ACM SIGPLAN Notices  (2016)

\bibitem{pek2014natural}
Pek, E., Qiu, X., Madhusudan, P.: Natural proofs for data structure
  manipulation in c using separation logic. In: ACM SIGPLAN Notices (2014)

\bibitem{piskac2008deciding}
Piskac, R., de~Moura, L., Bj{\o}rner, N.: Deciding effectively propositional
  logic with equality. Tech. rep., Technical Report MSR-TR-2008-181, Microsoft
  Research (2008)

\bibitem{pnueli2002small}
Pnueli, A., Rodeh, Y., Strichman, O., Siegel, M.: The small model property: How
  small can it be? Information and computation  (2002)

\bibitem{qiu2013natural}
Qiu, X., Garg, P., {\c{S}}tef{\u{a}}nescu, A., Madhusudan, P.: Natural proofs
  for structure, data, and separation. ACM SIGPLAN Notices  (2013)

\bibitem{naturalsynth17}
Qiu, X., Solar-Lezama, A.: Natural synthesis of provably-correct data-structure
  manipulations. Proc. ACM Program. Lang.  1(OOPSLA),  65:1--65:28 (Oct 2017),
  \url{http://doi.acm.org/10.1145/3133889}

\bibitem{rabin1969decidability}
Rabin, M.O.: Decidability of second-order theories and automata on infinite
  trees. Transactions of the american Mathematical Society  (1969)

\bibitem{hierarchical2005}
Sofronie-Stokkermans, V.: Hierarchic reasoning in local theory extensions. In:
  Nieuwenhuis, R. (ed.) Automated Deduction -- CADE-20. pp. 219--234. Springer
  Berlin Heidelberg, Berlin, Heidelberg (2005)

\bibitem{localtheory2013}
Sofronie-Stokkermans, V.: On combinations of local theory extensions. In:
  Programming Logics: Essays in Memory of Harald Ganzinger. pp. 392--413 (2013)

\bibitem{solar2006combinatorial}
Solar-Lezama, A., Tancau, L., Bodik, R., Seshia, S., Saraswat, V.:
  Combinatorial sketching for finite programs. ACM SIGOPS Operating Systems
  Review  (2006)

\end{thebibliography}

\newpage

\appendix

\section{Encoding \texorpdfstring{$M_\textsf{three}$}{Mthree} in \logic}
\label{app:mthree}

%!TEX root = main.tex

% \begin{align}
% \label{eqn:mthree1}
% M_\textsf{three}(n) = 
% \begin{cases}
% n - 30 & \text{ if } n > 13 \\
% M_\textsf{three}(M_\textsf{three}(M_\textsf{three}(n+61))) & \text{ otherwise }
% \end{cases}
% \end{align}

We are interested in synthesizing a straight line program
that implements the function $M_\textsf{three}$, and
can be expressed as a \texttt{term} over the grammar in Figure~\ref{fig:gram}.

Let us see how to encode this synthesis problem in \logicWithPeriod
First, let us fix the maximum height of the term we are looking for,
say to be 2.
Then, the program we want to synthesize can be represented 
as a tree of height at most $2$ such that every node in 
the tree can have $\leq 3$ child nodes 
(because the maximum arity of any function 
in the above grammar is $3$, corresponding to $\texttt{ite}$).
A skeleton of such a expression tree is shown in Figure~\ref{fig:skeleton}.
Every node in the tree is named according to its path from the root node.

The synthesis problem can then be encoded as the formula

% \vspace{-0.5cm}

\begin{align}
\begin{split}
\phi_{M_\textsf{three}} \equiv \, & (\exists n_0, n_{00}, n_{01}, \ldots n_{022} : \sigma_0) \; 
(\underbrace{\exists \lft, \md, \rgt : \sigma_0, \sigma_0}_{\textcolor{black}{\textbf{Existentially quantified relations}}}) \\
& (\exists \add,\ite,\lt,\eq,\gt,\inp, C_1, C_2, C_3 : \sigma_\lbl)\; \\
& (\exists {c_1, c_2, c_3 : \mathbb{N}}) \; (\underbrace{\exists f_\lbl : \sigma_0, \sigma_\lbl}_{\textcolor{black}{\textbf{Existentially quantified functions}}}) \\
& \quad\quad \varphi_\textsf{well-formed}\\
& \quad \quad \land (\forall x : \mathbb{N}) (\underbrace{\forall g_\val^0,g_\val^1,g_\val^2,g_\val^3 : \sigma_0, \mathbb{N}}_{\textcolor{black}{\textbf{Universally quantified functions}}}) \;  (\varphi_{\textsf{semantics}} \implies \varphi_{\textsf{spec}})
\end{split}
\end{align}
Here, the nodes are elements of the foreground sort $\sigma_0$.
The binary relations $\lft, \md, \rgt$ over the foreground sort
will be  used to assert that
a node $n$ is the left,middle, right child respectively of node $n'$ :
$\lft(n', n)$, $\md(n', n)$, $\rgt(n', n)$.
The operators or \emph{labels} for nodes belong to the background sort $\sigma_\lbl$,
and can be one of $\add \, (+)$, $\ite \, (\mathtt{ite})$, $\lt \, (< 0)$,
$\gt \, (> 0)$, $(\eq \, (= 0))$, $\inp$ (denoting the input to our program),
or constants $C_1, C_2, C_3$ (for which we will synthesize natural constants $c_1, c_2, c_3$
in the (infinite) background sort $\mathbb{N}$).
The function $f_\lbl$ assigns a label to every node in the program,
and the formula $\varphi_\textsf{well-formed}$ asserts some sanity conditions:
\begin{align}
\begin{split}
\varphi_\textsf{well-formed}  \equiv & \bigwedge\limits_{\rho \neq \rho'} n_\rho \neq n_{\rho'} \; 
\land \lft(n_0, n_{00}) \land \bigwedge\limits_{\rho \neq 00} \neg(\lft(n_0, n_\rho))) \land \cdots\\
& \land \neg (\add = \ite) \land \neg (\add = \lt) \land \cdots \land \neg(C_1 = C_3) \land \neg (C_2 = C_3) \\
& \land \bigwedge\limits_{\rho}   (f_\lbl(n_\rho) {=} \add) \lor (f_\lbl(n_\rho) {=} \ite) \lor \cdots \lor (f_\lbl(n_\rho) {=} C_3 )
\end{split}
\end{align}
% \vspace{-0.2cm}

The formula $\varphi_\textsf{semantics}$ asserts that the `meaning'
of the program can be inferred from the meaning of the components of the program.
The functions $g_\val^0, g_\val^1, g_\val^2, g_\val^3$, 
will assigns value to nodes from $\mathbb{N}$,
for this purpose :
\begin{align}
\begin{array}{rcl}
\varphi_\textsf{semantics} & \equiv & \varphi_\add \land \varphi_\ite \land \varphi_\lt \land \varphi_\eq \land \varphi_\gt \land \varphi_\inp \land \varphi_{C_1} \land \varphi_{C_2} \land \varphi_{C_3}
\end{array}
\end{align}

where each of the formulae $\varphi_\add, \cdots, \varphi_{C_3}$ specify the semantics of each node when labeled with these operations:

\begin{align}
\begin{array}{rcl}
\varphi_\add & \equiv \bigwedge\limits_{\rho, \rho_1, \rho_2} 
\big(f_\lbl(n_\rho) = \add \land \lft(n_\rho, n_{\rho_1})  \land \md(n_\rho, n_{\rho_2}) & \\
& \implies \bigwedge\limits_{i =0, 1,2, 3}
g_\val^i(n_\rho) = g_\val^i(n_{\rho_1}) + g_\val^i(n_{\rho_2}) \big) \\\end{array}
\end{align}

\begin{align}
\begin{array}{rcl}
\varphi_\ite & \equiv \bigwedge\limits_{\rho, \rho_1, \rho_2, \rho_3} 
\Big[f_\lbl(n_\rho) = \ite \land \lft(n_\rho, n_{\rho_1})  \land \md(n_\rho, n_{\rho_2}) \land \rgt(n_\rho, n_{\rho_3}) & \\
& \implies \bigwedge\limits_{i =0, 1,2, 3}
 \big( g_\val^i(n_{\rho_1}) = 1 \implies g_\val^i(n_{\rho}) = g_\val^i(n_{\rho_2}) \\
 &  \quad\quad\quad\quad\quad\quad \land \, g_\val^i(n_{\rho_1}) = 0 \implies g_\val^i(n_{\rho}) = g_\val^i(n_{\rho_3}) \big)\Big] 
 \end{array}
\end{align}

\begin{align}
\begin{array}{rcl}
\varphi_\lt & \equiv \bigwedge\limits_{\rho, \rho_1} 
\Big[f_\lbl(n_\rho) = \lt \land \lft(n_\rho, n_{\rho_1}) & \\
& \implies \bigwedge\limits_{i =0, 1,2, 3}
 \big( g_\val^i(n_{\rho_1}) < 0 \implies g_\val^i(n_{\rho}) = 1 \\
 &  \quad\quad\quad\quad\quad\quad \land \, g_\val^i(n_{\rho_1}) \geq 0 \implies g_\val^i(n_{\rho}) = 0 \big)\Big] 
 \end{array}
\end{align}

\begin{align}
\begin{array}{rcl}
\varphi_\eq & \equiv \bigwedge\limits_{\rho, \rho_1} 
\Big[f_\lbl(n_\rho) = \lt \land \lft(n_\rho, n_{\rho_1}) & \\
& \implies \bigwedge\limits_{i =0, 1,2, 3}
 \big( g_\val^i(n_{\rho_1}) = 0 \implies g_\val^i(n_{\rho}) = 1 \\
 &  \quad\quad\quad\quad\quad\quad \land \, g_\val^i(n_{\rho_1}) \neq 0 \implies g_\val^i(n_{\rho}) = 0 \big)\Big] 
 \end{array}
\end{align}

\begin{align}
\begin{array}{rcl}
\varphi_\gt & \equiv \bigwedge\limits_{\rho, \rho_1} 
\Big[f_\lbl(n_\rho) = \lt \land \lft(n_\rho, n_{\rho_1}) & \\
& \implies \bigwedge\limits_{i =0, 1,2, 3}
 \big( g_\val^i(n_{\rho_1}) > 0 \implies g_\val^i(n_{\rho}) = 1 \\
 &  \quad\quad\quad\quad\quad\quad \land \, g_\val^i(n_{\rho_1}) \leq 0 \implies g_\val^i(n_{\rho}) = 0 \big)\Big] 
 \end{array}
\end{align}\\

The formula $\varphi_\inp$ states that for a node
labeled $\inp$, the value of that node is the input to $M_\textsf{three}$.
Hence, such a node $n_\rho$ evaluates to $x$, $x+61$, $g_\val^1(n_0)$
and $g_\val^2(n_0)$ respectively under $g_\val^0$, $g_\val^1$, $g_\val^2$ and $g_\val^3$:
\begin{align}
\begin{array}{rcl}
\varphi_\inp & \equiv \bigwedge\limits_{\rho} 
\Big[ f_\lbl(n_\rho) = \inp & \implies \\
& & g_\val^0(n_{\rho}) = x \\
& & \land g_\val^1(n_{\rho}) = x + 61  \\
& & \land g_\val^2(n_{\rho}) = g_\val^1(n_0)  \\
& & \land g_\val^3(n_{\rho}) = g_\val^2(n_0) \\
&  \quad\quad\quad\quad\quad\quad \Big] &
 \end{array}
\end{align}

Finally we have the semantics of constant labels:
\begin{align}
\begin{array}{rl}
\varphi_{C_1} & \equiv \bigwedge\limits_{\rho} 
\Big[f_\lbl(n_\rho) = C_1  \implies \bigwedge\limits_{i =0, 1,2, 3} g_\val^i(n_{\rho}) = c_1 \Big] 
 \end{array}
\end{align}

The formulae $\varphi_{C_2}$ and $\varphi_{C_3}$ are similar and thus skipped.\\

% \vspace{-0.2cm}

Last, the formula $\varphi_\textsf{spec}$ expresses the specification of the program
as in Equation~\eqref{eqn:mthree}.

\begin{align}
\begin{array}{rl}
\varphi_\textsf{spec} & \equiv \big( x > 13 \implies g_\val^0(n_0) = x - 30 \big) \\
& \land \big(x \leq 13 \implies g_\val^0(n_0) = g_\val^3(n_0) \big)
\end{array}
\end{align}

% \end{document}

% \input{appendix}

\end{document}